\documentclass[12pt,onecolumn]{IEEEtran} % arxiv Mode
\IEEEoverridecommandlockouts                              % This command is only
\usepackage[top = 1.0in, bottom = 0.75in, left = 0.75in, right = 0.75in]{geometry}
\usepackage{times}
\usepackage{booktabs}
\usepackage{multirow}
\usepackage{rotating}
\usepackage{multicol}
\usepackage{dsfont}
\usepackage{amssymb}
\usepackage{float}
\usepackage{cite}
\usepackage{eurosym}
\usepackage{amsmath}
\usepackage{setspace}
\usepackage{hhline}
\usepackage{psfrag}
\usepackage{graphicx}
\usepackage{subfigure}
\usepackage{acronym}
\usepackage{flushend}

\usepackage{color}

\begin{document}

%\doublespacing

%% Acronym definitions

% power
\acrodef{hpfc}[HPFC]{Hourly Price Forward Curve}
\acrodef{pfc}[PFC]{Price Forward Curve}
\acrodef{hdd}[HDD]{heating degree day}
\acrodef{cdd}[CDD]{cooling degree day}
\acrodef{vpp}[VPP]{virtual power plants}
\acrodef{epex}[EPEX]{European Power Exchange}
\acrodef{eex}[EEX]{European Energy Exchange}
\acrodef{omel}[OMEL]{Operadora del Mercado Espa\~nol de Electricidad}
\acrodef{res}[RES]{renewable energy sources}
\acrodef{fit}[FIT]{feed-in tariff}
\acrodef{eeg}[EEG]{"Erneuerbare-Energien-Gesetz"}
\acrodef{seg}[SEG]{"Stromeinspeisungsgesetz"}
\acrodef{pv}[PV]{photovoltaic}
\acrodef{tso}[TSO]{transmission system operator}
\acrodef{tsos}[TSO]{transmission system operators}
\acrodef{otc}[OTC]{over-the-counter}
\acrodef{pshp}[PSHP]{pumped storage hydro plant}
\acrodef{pshps}[PSHP]{pumped storage hydro plants}
\acrodef{csp}[CSP]{concentrating solar power}

\acrodef{entsoe}[ENTSO-E]{European Network of Transmission System Operators for Electricity}

% math
\acrodef{ode}[ODE]{ordinary differential equation}
\acrodef{lp}[LP]{linear program}
\acrodef{qp}[QP]{quadratic program}
\acrodef{fft}[FFT]{Fast Fourier Transformation}

% statistics
\acrodef{mape}[MAPE]{mean absolute prediction error}
\acrodef{mae}[MAE]{mean absolute error}
\acrodef{mse}[MSE]{mean square error}
\acrodef{sma}[SMA]{simple moving average}
\acrodef{ewma}[EWMA]{exponentially weighted moving average}
\acrodef{ols}[OLS]{ordinary least squares}
\acrodef{lad}[LAD]{least absolute deviation}
\acrodef{ladlasso}[LAD-LASSO]{Least Absolute Deviation with Least Absolute Shrinkage and Selection Operator}
\acrodef{lasso}[LASSO]{Least Absolute Shrinkage and Selection Operator}
\acrodef{sde}[SDE]{stochastic differential equation}
\acrodef{mle}[MLE]{Maximum Likelihood Estimator}
\acrodef{aic}[AIC]{Akaike Information Criterion}
\acrodef{bic}[BIC]{Baysian Information Criterion}
\acrodef{ar}[AR]{autoregressive}
\acrodef{arx}[ARX]{autoregressive with external inputs}
\acrodef{arma}[ARMA]{autoregressive moving average}
\acrodef{arima}[ARIMA]{autoregressive integrated moving average}
\acrodef{wn}[WN]{white noise}

% Machine Learning
\acrodef{fl}[FL]{fuzzy logic}
\acrodef{svm}[SVM]{support vector machine}
\acrodef{lssvm}[LSSVM]{least squares support vector machine}
\acrodef{elm}[ELM]{extreme learning machine}

% risk
\acrodef{var}[VaR]{Value-at-Risk}
\acrodef{cvar}[CVaR]{conditional Value-at-Risk}
\acrodef{par}[PaR]{Profit-at-Risk}
\acrodef{cpar}[CPaR]{conditional Profit-at-Risk}
\acrodef{nn}[NN]{neuronal networks}
\acrodef{ann}[ANN]{artificial neural networks}

%economics
\acrodef{gdp}[GDP]{gross domestic product}
\acrodef{oecd}[OECD]{Economic Co-operation and Development}
\acrodef{npv}[NPV]{net present value}
\acrodef{moc}[MOC]{Merit-Order-Curve}

% departments and companies
\acrodef{kti}[KTI]{Swiss Innovation Promotion Agency}
\acrodef{sqg}[SQG]{\textrm{swissQuant} Group AG}
\acrodef{eu}[EU]{European Union}

%\title{Are Energy-Only Power Markets a Failure?}
%\title{Revisiting the Merit-Order Effect \\ of Renewable Energy Sources (in Spot Markets)}
\title{Revisiting the Merit-Order Effect \\ of Renewable Energy Sources}
%\author{
%\IEEEauthorblockN{Marcus Hildmann, Andreas Ulbig, G\"oran Andersson}
%\IEEEauthorblockA{Power Systems Laboratory, Department of Electrical Engineering, ETH Zurich\\
%8092 Zurich, Switzerland\\
%{\textmd \{hildmann $|$ ulbig $|$ andersson\}@eeh.ee.ethz.ch}}\vspace{0.25cm}
%\IEEEauthorblockN{\textbf{\texttt{\textsc{Working Paper - Version Date 31 May 2013}}}}
%\IEEEauthorblockN{Andreas Ulbig}
%\IEEEauthorblockA{Power Systems Laboratory, ETH Zurich, Switzerland\\
%Email: ulbig@eeh.ee.ethz.ch}\vspace{0.1cm}
%Power Systems Laboratory, ETH Zurich, Switzerland%\and
%\IEEEauthorblockN{Marcelo Ito Parada}
%\IEEEauthorblockA{Power Systems Laboratory, ETH Zurich, Switzerland\\
%Email: parada@student.ethz.ch}
%\IEEEauthorblockN{G\"oran Andersson}
%\IEEEauthorblockA{Power Systems Laboratory, ETH Zurich, Switzerland\\
%Email: andersson@eeh.ee.ethz.ch}
%}

%\author{Marcus~Hildmann,~\IEEEmembership{Student Member,~IEEE},
%        Andreas Ulbig,~\IEEEmembership{Student Member,~IEEE}
%        and G\"oran Andersson,~\IEEEmembership{Fellow,~IEEE}% <-this % stops a space
\author{Marcus~Hildmann, Andreas Ulbig and G\"oran Andersson% <-this % stops a space
\thanks{M. Hildmann, A. Ulbig and G. Andersson are with the Power Systems Laboratory, Dept. of Information Technology and Electrical Engineering, ETH Zurich (Swiss Federal Institute of Technology), %Switzerland. \newline Email: \{hildmann, ulbig, andersson\}@eeh.ee.ethz.ch.}% <-this % stops a space
Switzerland. \newline Email: {\textmd hildmann$\;$\textbar$\;$ulbig$\;$\textbar$\;$andersson$\,$@$\,$eeh.ee.ethz.ch}}% <-this % stops a space 	
%}% <-this % stops a space
%\thanks{Working Paper - Version Date 31 May 2013}}
%\thanks{\textbf{Working Paper -- Version Date 23 December 2013} (Content change since Version of 1 July 2013: data updates and more indepth analysis of Fig.XYZ, text update of subsection %\ref{sec:MarginalCosts}.)}
}

%\markboth{10$^{\textrm{\MakeLowercase{th}}}$ International Conference on the European Energy Market (EEM13), Stockholm Sweden 2013}%
%{Shell \MakeLowercase{\textit{et al.}}: Bare Demo of IEEEtran.cls for Journals}

\maketitle
\thispagestyle{empty}

\begin{abstract}
An on-going debate in the energy economics and power market community has raised the question if energy-only power markets are increasingly failing due to growing feed-in shares from subsidized \acf{res}. The short answer to this is: No, they are not failing.
Energy-based power markets are, however, facing several market distortions, namely from the gap between the electricity volume traded at day-ahead markets versus the overall electricity consumption as well as the (wrong) regulatory assumption that variable \ac{res} generation,~i.e., wind and \ac{pv}, truly have zero marginal operation costs. 

In this paper we show that both effects over-amplify the well-known merit-order effect of \ac{res} power feed-in beyond a level that is explainable by underlying physical realities,~i.e., thermal power plants being willing to accept negative electricity prices to be able to stay online due to considerations of wear \& tear and start-stop constraints.
We analyze the impacts of wind and \ac{pv} power feed-in on the day-ahead market for a region that is already today experiencing significant \ac{fit}-subsidized \ac{res} power feed-in, the \acs{epex} German-Austrian market zone~($\approx\,$20\% \ac{fit} share).
 
%We show a comparison of the FIT-subsidized \ac{res} energy production volume to the day-ahead market volume and the overall load demand. Furthermore, a day-ahead market analysis based on the assumption that \ac{res} units have to feed-in with their assumed true marginal costs,~i.e., operation, maintenance and balancing costs, is performed. 
Our analysis shows that, if the necessary regulatory adaptations are taken,~i.e., increasing the day-ahead market's share of overall load demand and using the true marginal costs of \ac{res} units in the merit-order, energy-based power markets can remain functional despite high \ac{res} power feed-in. 
\end{abstract}

%\begin{IEEEkeywords}
%Electricity load forecasting, Electricity price modeling, Hourly Price Forward Curve, contract valuation
%\end{IEEEkeywords}

% reset acronym counter from abstract
\acresetall

% \fi
\IEEEpeerreviewmaketitle

\acresetall
\section{Introduction}\label{sec:introduction}

\begin{figure*}
	\centering
	%\vspace{-0.3cm}
     %\centerline{\psfig{figure=figures/ComparisonPeakShapeYearly.eps,width=\linewidth} }
    \psfrag{x1}[cc][cc]{\scriptsize\shortstack{Year}}
    \psfrag{y1}[cc][cc]{\scriptsize\shortstack{GW}}
    \includegraphics[width=0.475\textwidth]{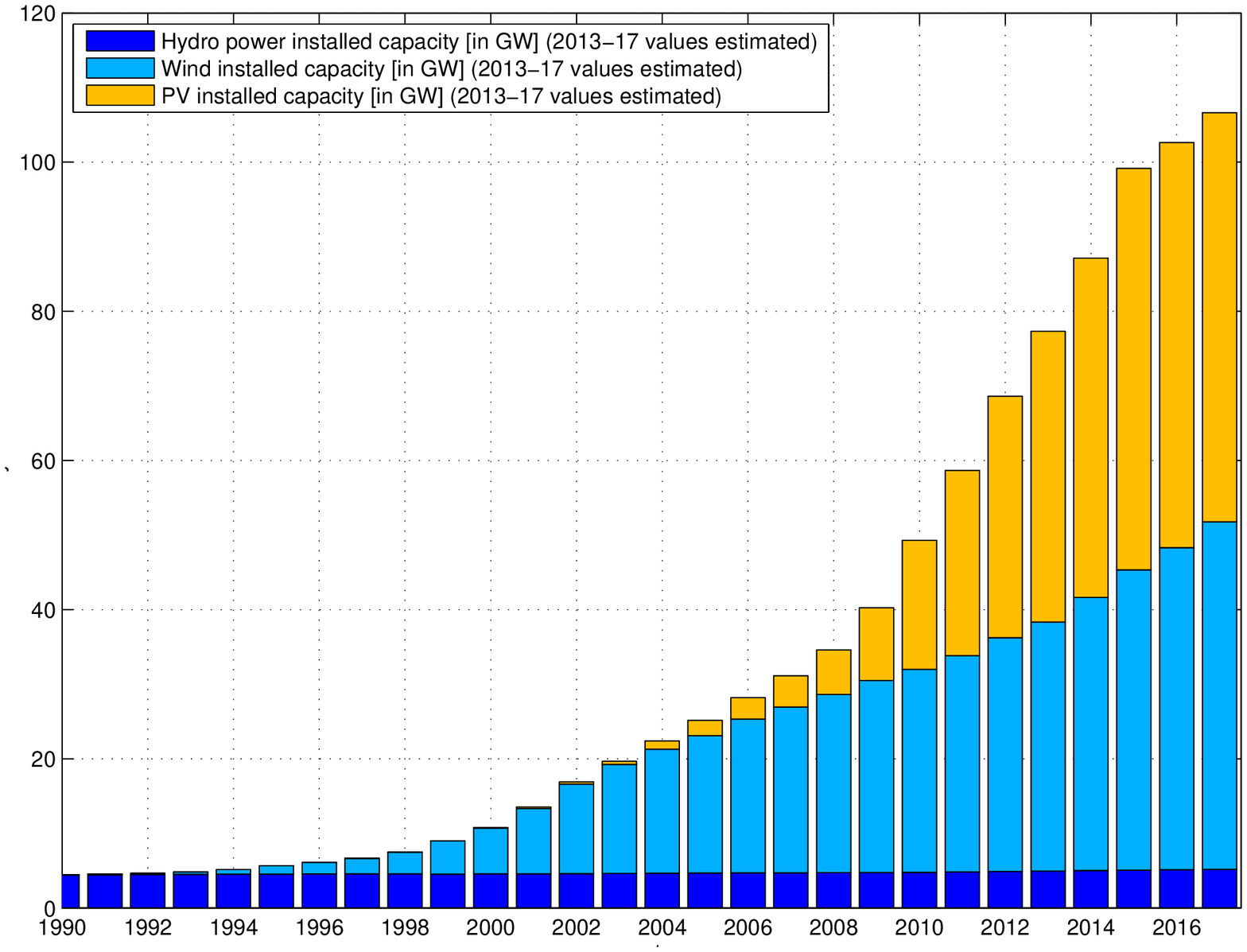}
    %\centerline{\psfig{figure=matlab/figures/modifiedAskCurve, width=1\linewidth} }
    \vspace{-0.20cm}
    %\vspace{-0.5cm}
	%\vspace{-0.3cm}
    %\centerline{\psfig{figure=figures/ComparisonPeakShapeYearly.eps,width=\linewidth} }
    \psfrag{x1}[cc][cc]{\scriptsize\shortstack{Year}}
    \psfrag{y1}[cc][cc]{\scriptsize\shortstack{TWh}}
    \includegraphics[width=0.475\textwidth]{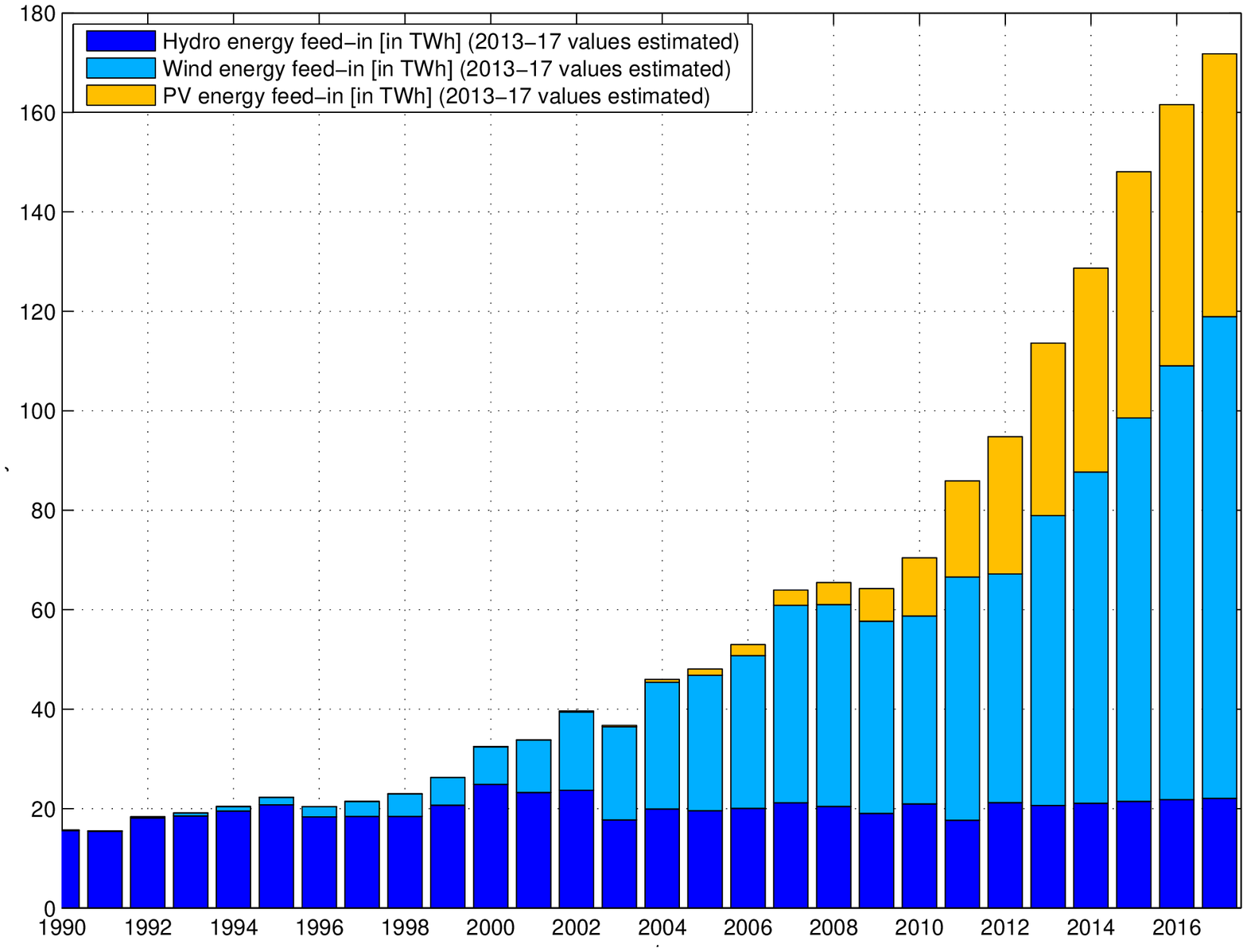}
    %\centerline{\psfig{figure=matlab/figures/modifiedAskCurve, width=1\linewidth} }
 	%\vspace{-0.10cm}
    \caption{Overview of \acf{res} deployment in Germany (1990--2017). Values for 2013--2017 are best estimates.\newline a)~Evolution of installed power capacity of wind turbine and \ac{pv} units. b)~Evolution of annual feed-in of wind turbine and \ac{pv} units.}
    \label{fig:RES_Overview}
    %\vspace{-0.3cm}
\end{figure*}

Around the same time as the liberalization of many of the European electricity markets in the early 1990s, government support schemes with the specific goal of promoting large-scale deployment of \acf{res} were introduced. The German Renewable Energy Act, \ac{eeg}, a well-known support scheme, provides a favorable \ac{fit} for a variety of \ac{res} since the year~2000, building on its predecessor, the \emph{Stromeinspeisungsgesetz}~(1990). It gives priority to electric power feed-in from \ac{res} over power feed-in from conventional power plants,~i.e., fossil- and nuclear-fueled thermal and old, often large hydro-based power plants. This favorable investment case has generated installed capacities of about 35~GW each for wind and \ac{pv} units by year-end~2013~\cite{REN21:2014}. The evolution of installed variable \ac{res} capacities and annual energy production in Germany from 1990 to 2012 \cite{AGEB2013}, including an outlook till~2017 \cite{BDEW2013}. 
%The original goal of \acp{fit},~i.e., large-scale \ac{res} deployment (\figurename~\ref{fig:installedCapacity}) and significant RES energy shares (\figurename~\ref{fig:overallProduction2012}) is achieved.
The original goal of \acp{fit},~i.e., large-scale \ac{res} deployment (\figurename~\ref{fig:RES_Overview}a) and significant RES energy shares (\figurename~\ref{fig:RES_Overview}b) is achieved.

%On the other hand, the provision of independent payments from \ac{fit} for 20 years results in high costs for the electricity consumers. %The so called "EEG Umlage" is 0.036 \euro/kWh in 2012 and will rise to 0.053 \euro/kWh in 2013, which reflects roughly 20\% of the end consumer tariff.
With a combined installed capacity of wind \& \ac{pv} units of around 70~GW by year-end~2013, somewhat higher than the average load demand in Germany (63--68~GW dependent on load demand measure~\cite{IEA_stats,EntsoePortal2012}), wind \& \ac{pv} units clearly cannot be treated as exotic, marginal electricity sources anymore. The current \ac{res} production has already significant effects on the power market, notably in the form of the so-called \emph{merit-order effect}. Especially the decoupling of day-ahead market prices and \ac{res} feed-in due to \ac{fit} regulations, results in lower average day-ahead price levels and also in negative day-ahead prices for several hours each month. 
In today's European power market environment in which significant over-capacities exist~\cite{Economist:2014}, yearly average spot market prices as well as base-peak spreads in Europe are at their lowest in years~\cite{EPEX_reports}. One effect of this is that flexible power plants such as gas-fired units cannot be operated profitably because peak day-ahead prices are too often below their marginal operation costs. Another effect is that due to the also decreasing spread of peak/base day-ahead prices, the profitability of operating energy storage facilities,~i.e., \ac{pshps} has been diminishing in recent years~\cite{Hildmann2011a}.

While in the long run support schemes for \ac{res} units are likely to be phased-out, and they will thus have to compete in a normal market setup, in the mid-term the power market structure may have to be adapted to the increasing effects of \ac{res} deployment. 
For the subsequent investigation the two relevant perspectives have to be taken into account:
\begin{enumerate}
\item \emph{Producer perspective}: How to keep the profitability of necessary dispatchable base and peak load power plants in a power market with high shares of \ac{res}?
\item \emph{Consumer perspective}: How to lower the cost of \ac{fit}~schemes and enable the transition to a less subsidized (and eventually true) power market?
\end{enumerate}

%\begin{figure}[h]
%	\centering
%	%\vspace{-0.3cm}
%     %\centerline{\psfig{figure=figures/ComparisonPeakShapeYearly.eps,width=\linewidth} }
%    \psfrag{x1}[cc][cc]{\scriptsize\shortstack{Year}}
%    \psfrag{y1}[cc][cc]{\scriptsize\shortstack{GW}}
%    \includegraphics[width=0.475\textwidth]{installedCapacity.eps}
%    %\centerline{\psfig{figure=matlab/figures/modifiedAskCurve, width=1\linewidth} }
%    \vspace{-0.20cm}
%    \caption{Evolution of installed capacity of wind and \ac{pv} units in Germany.}
%    \label{fig:installedCapacity}
%    %\vspace{-0.5cm}
%\end{figure}
%
%\begin{figure}[h]
%	\centering
%	%\vspace{-0.3cm}
%    %\centerline{\psfig{figure=figures/ComparisonPeakShapeYearly.eps,width=\linewidth} }
%    \psfrag{x1}[cc][cc]{\scriptsize\shortstack{Year}}
%    \psfrag{y1}[cc][cc]{\scriptsize\shortstack{TWh}}
%    \includegraphics[width=0.475\textwidth]{overallProduction2012.eps}
%    %\centerline{\psfig{figure=matlab/figures/modifiedAskCurve, width=1\linewidth} }
% 	\vspace{-0.20cm}
%    \caption{Evolution of annual feed-in of wind and \ac{pv} units in Germany.}
%    \label{fig:overallProduction2012}
%    %\vspace{-0.3cm}
%\end{figure}

\subsection{Producer Perspective}
In Germany, and similarly in Austria, all \ac{fit}-supported \ac{res} power feed-in is settled at the day-ahead market of the \ac{epex} either directly, e.g.~in the form of power supply bids, or indirectly, e.g.~in the form of reduced load demand bids. This construction results in two major effects on the day-ahead market. Firstly, because of the decoupling of \ac{fit}-supported \ac{res} feed-in from day-ahead spot prices, the average spot price decreases, since all those bids will be located on the negative side of the supply curve, e.g.~going as low as $-3000$~\euro/MWh, in order to guarantee the settlement. Besides the generally lower day-ahead spot price levels, negative day-ahead prices tend to occur, especially in \emph{low demand \& high wind} situations. Secondly, because of the large deployment of \ac{pv}, the peak/base spread is decreasing significantly~\cite{Hildmann2011a,EPEX_reports}.
Combined with today's significant over-capacities of conventional power plants in Europe~\cite{Economist:2014}, these developments result in profitability problems of both base and peak load power plants as well as short-term storage units (i.e.~\acp{pshp}). 

Short- and mid-term, a completely \ac{res} based electricity production on its own can neither provide the necessary production volume nor the necessary reliability for fulfilling the load demand at all times throughout the year. Conventional power plants, whose generation output can be controlled, are still necessary for providing base and peak load demand, buffer the lack of wind and \ac{pv} power feed-in during absences of wind \& sunshine and, generally, cover the prediction error of variable \ac{res} power feed-in. 

One proposed solution for this problem is the implementation of capacity markets, where (conventional) power plants are rewarded not only for energy delivery but also for providing a firm power capacity~\cite{DENA2012}. 
Capacity markets help to reduce power plant investment risk as they provide an additional revenue stream for power plant owners. 
However, the introduction of capacity markets would add yet another subsidy mechanism, preventing both an efficient functioning of the energy-only power market and, in the long run, an affordable energy transition to significant \ac{res} shares in the electricity sector. The idea of introducing capacity markets is thus challenged by others, who argue that energy-only power markets can actually function efficiently for power systems with even higher RES shares than today~\cite{Nicolosi:2012, Nicolosi:2014}.
%While the majority of \ac{res} are wind and \ac{pv}, the delivered volume is uncertain and depends on several seasonal in different frequencies. %

\subsection{Consumer Perspective}
Out of the many possible support schemes for \ac{res}, \ac{fit}-based schemes turned out to be highly effective energy policy instruments. As a result, large-scale deployment of \ac{res}, especially wind and \ac{pv} has been remarkably successful. The actual \ac{fit} payments,~i.e., the difference between \ac{fit} tariffs and hourly day-ahead market prices, is financed in Germany via an additional levy on electricity consumption~(\emph{EEG Umlage}). 
Due to the EEG \ac{fit} scheme's great success, this levy has become significant, i.e.~6.24~\euro ct/$\textrm{kWh}_\textrm{el}$, or about 22\%, of an average residential electricity rate of 29~\euro ct/$\textrm{kWh}_\textrm{el}$ (year~2014). 
The average German retail consumer electricity price is nowadays a midst the highest in Europe~\cite{EUROSTAT:2014} whereas the average day-ahead whole-sale electricity price is one of the lowest~\cite{EPEX_reports}. To prevent additional long term price increases, the consumer has the interest that costs for \ac{fit} schemes are lowered and that the integration of \ac{res} power feed-in into competitive power market frameworks is achieved.

\subsection{Outline}
While in the long run, subsidy schemes for \ac{res} units will eventually be phased-out, the uncertainty of \ac{res} electricity production,~i.e., the inherent mismatch between production forecasts and actual production, will persist. As an intermediate step to a fully competing power market setup for \ac{res}, where they face a price risk and also have to pay for the forecast error, corrections of existing power market frameworks are necessary. Such corrections have to ensure that consumer and producer perspectives are satisfied, while minimizing regulatory actions.

In this paper we analyze the following two aspects of the \ac{epex} power market, while leaving the currently existing day-ahead power market setup intact:
\begin{enumerate}
\item Comparison of hourly power feed-in volume of wind \& \ac{pv} units to the overall load demand volume.
\item Estimation of resulting day-ahead market price levels in case that the given wind \& \ac{pv} electricity volumes are settled for a range of assumed possible marginal operation costs (0--20~\euro/$\textrm{MWh}_\textrm{el}$), see Section \ref{sec:marCostLiterature} for the reasoning behind non-zero marginal costs of wind \& \ac{pv} units and an indicative literature review.
\end{enumerate}
Based on those assumptions we will discuss the effect on the power market as well as on power plants. We will finally elude on the original question if substantial alterations to power market setups such as the introduction of capacity markets are really necessary or whether there are simpler ways to reduce market distortions in energy-only markets caused by large amounts of \ac{res} feed-in.

\section{Data}

The analysis is based on high-resolution time-series data of the German power system, provided by the transparency platform of the \ac{eex} \cite{EEXTransparency} and the \ac{entsoe} data portal~\cite{EntsoePortal2012}. The analysis is done for the full years 2011--13. Table~\ref{tab:dataSources} lists the employed time-series data and data sources.
\begin{table}[h]
\begin{center}
\renewcommand{\arraystretch}{1.0}
\centering

\footnotesize
\caption{used data and sources}

\label{tab:dataSources}

\begin{tabular}{lc}

%\toprule
%\hline
\toprule
\textbf{Time-Series Data} & \textbf{Sources} \\
%\hhline{|=====|}
\toprule
%\midrule
%\hline
Vertical grid load & \ac{entsoe}\\
\midrule
German day-ahead price & \ac{eex}\\
\midrule
German day-ahead bid/ask curve  & \ac{eex}\\
\midrule
German wind feed-in  & \ac{eex} \\
\midrule
German \ac{pv} feed-in & \ac{eex} \\
%Reflects prices in neighboring markets when no & \multirow{2}{*}{should}\\
%local exchange-traded market prices exist &\\
%%Correctly extends neighboring Futures when this data lacks & should \\
%\midrule
%Prefer Future products with higher frequency or  & \multirow{2}{*}{should}\\
%liquidity & \\
\bottomrule
%\hline
\end{tabular}
\end{center}
\end{table}

Vertical grid load, wind and \ac{pv} feed-in are available with 15-min. resolution, while day-ahead prices and bid/ask curves are available with hourly resolution. All analysis with data of both resolutions are done on hourly resolutions and the 15-min. resolution data is down-sampled to fit hourly resolution.
Vertical grid load is adjusted to match the official total electricity supplied in Germany, i.e.~561.3~TWh in 2011 (566.7~TWh in 2012)~\cite{AGEB2013,IEA_stats}. The bid/ask curve data as provided by \ac{eex} seemed to be somewhat filtered and restructured (exactly 200 demand/supply bids for each hourly step); supposedly to blur the exact bid and ask structure. The available data quality is, however, fully sufficient for our analysis. (Daylight saving time effects were appropriately removed.)

\section{Quantitative Analysis of EPEX Spot}\label{sec:QuantAnalysis}

\subsection{Spot Market Volume and \ac{res} Production}
The \acf{epex} is the dominant day-ahead and intra-day power market in Central and Western Europe, directly serving Austria, France, Germany and Switzerland. Here, most of the day-ahead spot market trading is performed via~\ac{epex}. The yearly clearing volume of the day-ahead spot market of \ac{epex}'s combined German/Austrian market zone amounted to around $224.6\,\textrm{TWh}_\textrm{el}$ in 2011~\cite{epex}. However, this is only about $40.0\,\%$ (or $35.8\,\%$) of the total electricity supplied of around $561.3\,\textrm{TWh}_\textrm{el}$ in Germany (or $\approx 627.8\,\textrm{TWh}_\textrm{el}$ in Germany and Austria combined) for~2011. By~2013, these market shares had risen to $44.2\,\%$~(DE) and $39.2\,\%$~(DE+AT), respectively.
All of Germany's \ac{fit}-subsidized \ac{res} generated electricity has to be traded on the \ac{epex} day-ahead market (DE+AT market zone). In~2011 this amounted to around $101.3\,\textrm{TWh}_\textrm{el}$ or 16.8\% of total electricity consumption. Out of the overall \ac{fit}-subsidized \ac{res} feed-in $18.4\,\%$ came from \ac{pv} and $45.1\,\%$ from wind turbines~\cite{AGEB2013} (and $\approx 108\,\textrm{TWh}_\textrm{el}$ combined for Germany and Austria). Variable \ac{res} units thus contributed about $63.5\,\%$ of all \ac{fit}-subsidized renewable electricity production. By~2013, the \ac{fit}-subsidized \ac{res} share had risen to $112\,\textrm{TWh}_\textrm{el}$~(DE) and $119\,\textrm{TWh}_\textrm{el}$~(DE+AT), respectively.

\subsection{Time-Series Analysis}

In the following, a thorough analysis of the respective yearly time-series of the \ac{epex} cleared trading volume (German/Austrian market zone), the \ac{fit}-subsidized renewable electricity production and the final electric load demand, both for Germany only, is performed. In the remainder of this section, all result figures were calculated using hourly power feed-in and load demand time-series for the full-year~2011~(8'760 hours). The results for the years~2012 and~2013 are qualitatively similar, exhibiting a steady rise of both \ac{res} energy as well day-ahead market shares with respect to total load demand.

%% Metric A : EPEX volume / ENTSO-E volume
%\vspace{-0.20cm}
\subsubsection{EPEX Trading Volume versus Load Demand}

\begin{figure}[tb]
    \centering
    %\centerline{\psfig{figure=figures/ComparisonPeakShapeYearly.eps,width=\linewidth} }
    \psfrag{x}[cc][cc]{\scriptsize\shortstack{Relative volume ([0 ... 1] = [0\% ... 100\%])}}
    \psfrag{y}[cc][cc]{\scriptsize\shortstack{Hours per year}}
    \includegraphics[width=0.55\textwidth]{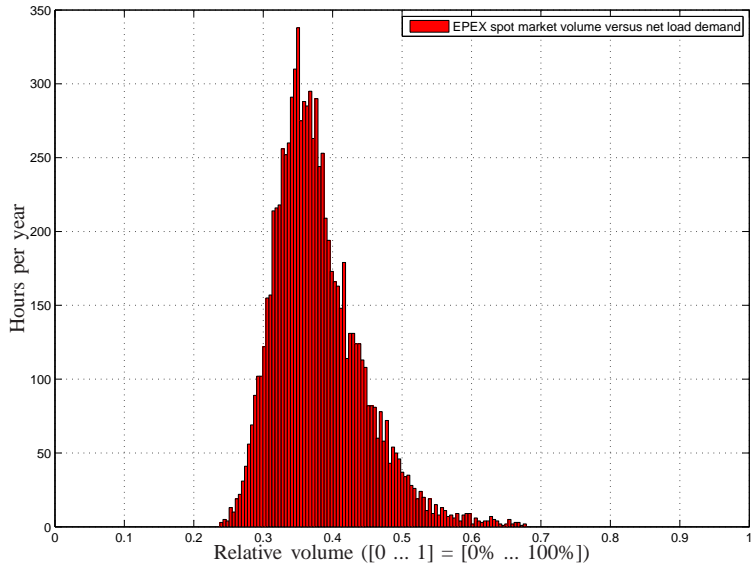}
    %\centerline{\psfig{figure=matlab/figures/modifiedAskCurve, width=1\linewidth} }
    \caption{Histogram of EPEX day-ahead spot volume versus German net load demand volume~(in~2011).}
    \label{fig:MetricA}
    \vspace{-0.2cm}
\end{figure}

The amount of the cleared day-ahead market volume as a function of final electricity consumption for a given hour oscillates between a minimum contribution of $23.8\,\%$ and a maximum contribution of $67.9\,\%$. A histogram of the \ac{epex} day-ahead volume versus German total load demand volume in~2011 is shown in~\figurename~\ref{fig:MetricA}. The average ratio between cleared day-ahead market volume and electricity demand over the course of the year was $35.8\,\%$~(volume-weighted mean). It contributed thus a bit less than half,~i.e., two-fifth, of all electricity demand in the German/Austrian \ac{epex} market zone for the full-year~2011.
%
%Metric A Mean [min/max]: 0.379 [0.238 / 0.679]
%Metric A Mean vol-based: 0.373
%Metric A Median: 0.368
%Yearly Load Demand: 602600000
%Yearly EPEX Volume: 224636960

%%  Metric B : EEG volume / ENTSO-E volume
%\vspace{-0.20cm}
\subsubsection{RES feed-in versus Load Demand}

The amount of \ac{fit}-subsidized \ac{res} production as a function of total electricity consumption for a given hour oscillates between a minimum contribution of $5.1\,\%$ and a maximum contribution of $49.1\,\%$. The mean contribution over the full year is $15.6\,\%$~(volume-weighted mean). The \ac{pv} energy share of net load demand varies between zero and $22.2\,\%$  with a volume-weighted mean of $3.4\,\%$ (in~2011). The wind energy share of net load demand varies between zero and $39.3\,\%$  with a volume-weighted mean of $8.7\,\%$ (in~2011).
%
%Metric B (EEG) Mean [min/max]: 0.170 [0.051 / 0.491]
%Metric B (EEG) Mean vol-based: 0.156
%Metric B (varRES) Mean [min/max]: 0.107 [0.000 / 0.417]
%Metric B (\ac{pv}) Mean [min/max]: 0.029 [0.000 / 0.222]
%Metric B (\ac{pv}) Mean vol-based: 0.029
%Metric B (Wind) Mean [min/max]: 0.078 [0.000 / 0.393]
%Metric B (Wind) Mean vol-based: 0.070
%Metric B (otherEEG) Mean vol-based: 0.057
%Metric B median (EEG): 0.155
%Metric B median (\ac{pv}): 0.001
%Metric B median (Wind): 0.056
%Yearly Load Demand: 650295232
%Yearly EPEX Volume: 224636960
%Yearly EEG Volume: 101134881

%%  Metric C : EEG volume / EPEX volume
%\vspace{-0.20cm}
\subsubsection{RES feed-in versus EPEX Trading Volume}
\begin{figure}[tb]
    \centering
    %\centerline{\psfig{figure=figures/ComparisonPeakShapeYearly.eps,width=\linewidth} }
    \psfrag{x}[cc][cc]{\scriptsize\shortstack{Relative volume ([0 ... 1] = [0\% ... 100\%])}}
    \psfrag{y}[cc][cc]{\scriptsize\shortstack{Hours per year}}
    \includegraphics[width=0.55\textwidth]{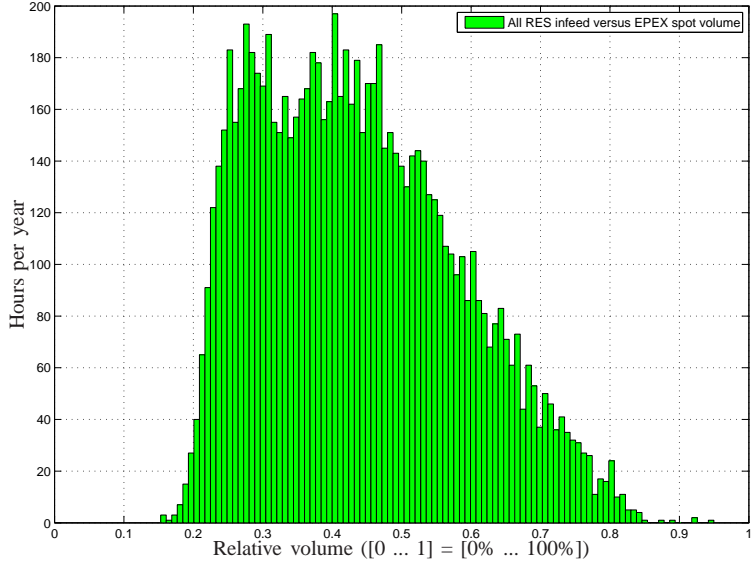}
    %\centerline{\psfig{figure=matlab/figures/modifiedAskCurve, width=1\linewidth} }
    \vspace{-0.1cm}
    \caption{Histogram of subsidized \ac{res} feed-in volume versus EPEX day-ahead spot volume~(in~2011).}
    \label{fig:MetricC}
    \vspace{-0.4cm}
\end{figure}

The amount of \ac{fit}-subsidized \ac{res} production that is bid into the spot market as a function of overall cleared day-ahead market volume for a given hour varies significantly over the course of the year. The ratio varies between a minimum of $15.3\,\%$ and a maximum of $95.0\,\%$. The mean contribution of \ac{fit}-subsidized \ac{res} feed-in to the total day-ahead market volume over the full year is $45.0\,\%$  (volume-weighted mean). \ac{fit}-subsidized \ac{res} generated electricity contributes thus a bit less than half of all electricity generation that is traded in the German/Austrian \ac{epex} market zone. A histogram of the \ac{fit}-subsidized \ac{res} feed-in volume versus the \ac{epex} day-ahead volume in 2011 is given in~\figurename~\ref{fig:MetricC}.
The \ac{pv} energy share of the day-ahead market volume varies between zero and $47.4\,\%$  with a volume-weighted mean of $8.6\,\%$ (in~2011). In comparison to this, the wind energy share of the day-ahead market volume varies between zero and $67.5\,\%$  with a volume-weighted mean of $21.8\,\%$ (in~2011). 
Other EEG feed-in,~i.e., biomass, landfill and bio gas as well as small hydro, make up a volume-weighted mean of $\approx\,20\,\%$ of total \ac{epex} volume (in~2011).
%
%Metric C Mean (EEG): 0.450
%Metric C Mean (\ac{pv}): 0.083
%Metric C Mean (Wind): 0.203
%Metric C (EEG) Mean [min/max]: 0.437 [0.153 / 0.950]
%Metric C (EEG) Mean vol-based: 0.450
%Metric C (\ac{pv}) Mean [min/max]: 0.075 [0.000 / 0.474]
%Metric C (\ac{pv}) Mean vol-based: 0.083
%Metric C (Wind) Mean [min/max]: 0.193 [0.000 / 0.675]
%Metric C (Wind) Mean vol-based: 0.203
%Metric C (other EEG) Mean vol-based: 0.164
%Metric C median (EEG): 0.422
%Metric C median (\ac{pv}): 0.002
%Metric C median (Wind): 0.157
\vspace{-0.1cm}

%\vspace{-0.20cm}
\subsection{Results}
\begin{figure}[tb]
    \centering
    %\centerline{\psfig{figure=figures/ComparisonPeakShapeYearly.eps,width=\linewidth} }
    \psfrag{x}[cc][cc]{\scriptsize\shortstack{Relative volume ([0 ... 1] = [0\% ... 100\%])}}
    \psfrag{y}[cc][cc]{\scriptsize\shortstack{Hours per year}}
    \includegraphics[width=0.55\textwidth]{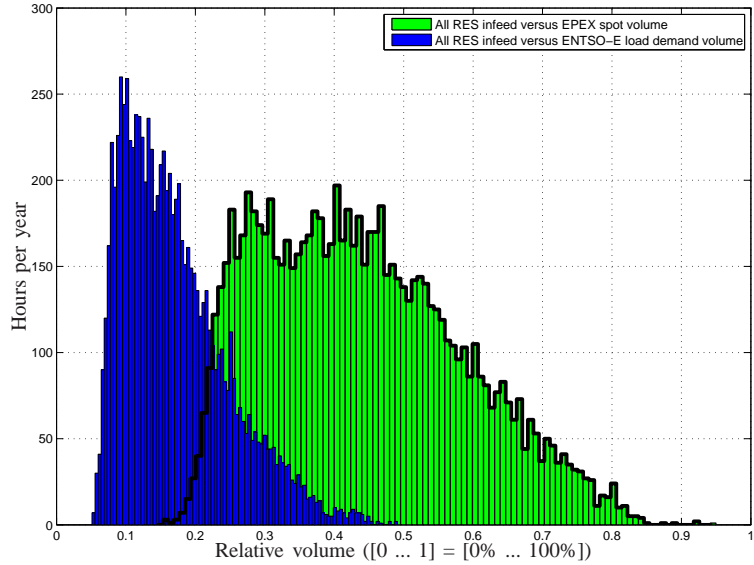}
    %\centerline{\psfig{figure=matlab/figures/modifiedAskCurve, width=1\linewidth} }
    \vspace{-0.1cm}
    \caption{Histograms of subsidized \ac{res} feed-in versus German net load demand volume and \ac{epex} day-ahead volume~(in~2011).}
    \label{fig:MetricBC}
    \vspace{-0.3cm}
\end{figure}

Although all \ac{fit}-subsidized \ac{res} generated electricity contributed less than a fifth of Germany's and Austria's combined total load demand ($\approx\,19.2\,\%$ in~2012, $\approx\,20.0\,\%$ in~2013), it constituted close to half ($\approx\,38.7\,\%$ in~2012, $\approx\,39.2\,\%$ in~2013) of the traded electricity volume of \ac{epex}'s German/Austrian market zone. This is clearly illustrated by the \emph{gap} between the histograms of both the \ac{fit}-subsidized \ac{res} feed-in volume versus total load demand and \ac{epex} day-ahead volume, respectively, as given in~\figurename~\ref{fig:MetricBC}.

The underlying reason for the observed volatile day-ahead price dynamics caused by the merit-order effect thus becomes obvious: The day-ahead market is, volume-wise, largely dominated by subsidized \ac{res} feed-in, which thus drives the dynamics of the day-ahead price clearing mechanism. This is in stark contrast to the physical situation in the German power system, in which \ac{fit}-subsidized \ac{res} feed-in plays a minor, but rising role. An obvious approach for mitigating highly fluctuating day-ahead prices within the existing energy-only power market is thus to increase the trading volume of day-ahead markets by incorporating more of the conventional electricity production that is currently sold bilaterally,~i.e.,~\emph{over-the-counter}~(OTC), outside the \ac{epex} day-ahead market.

\subsection{Comparison of EPEX and Nord Pool Spot Market} \label{sec:Nordpool}
Comparing the situation of the EPEX day-ahead market to the situation in a neighboring power market reveals interesting insights. The \emph{Nord Pool} is a mature multinational day-ahead spot market that comprises the north and north-eastern European countries Finland, Denmark, Norway, Sweden and Estonia \cite{Nordpool}. The yearly total load demand of the \emph{Nord Pool} countries ($\approx 440\,\textrm{TWh}_\textrm{el}$) is comparable in scale to the German/Austrian \ac{epex} market zone ($\approx 665\,\textrm{TWh}_\textrm{el}$) \cite{IEA_stats}. Both power markets comprise day-ahead and intra-day markets that are organized similarly (energy-only, with a strategic reserve in Scandinavia). \emph{Nord Pool} is, as a directly neighboring and well-connected market with a similarly high \ac{res} integration and a similarly small number of few big players with conventional plants, in fact an ideal candidate for a comparison.

\begin{figure}[b]
	\vspace{-0.40cm}
    \centering
    %\centerline{\psfig{figure=figures/ComparisonPeakShapeYearly.eps,width=\linewidth} }
    \psfrag{x}[cc][cc]{\scriptsize\shortstack{Years}}
    \psfrag{y}[cc][cc]{\scriptsize\shortstack{Market share [in \%]}}
	% Journal Version
    %\includegraphics[width=0.45\textwidth]{SpotVolume_RES_Comparison_DEAT.eps}
	% Arxiv-Version	
	\includegraphics[width=0.70\textwidth]{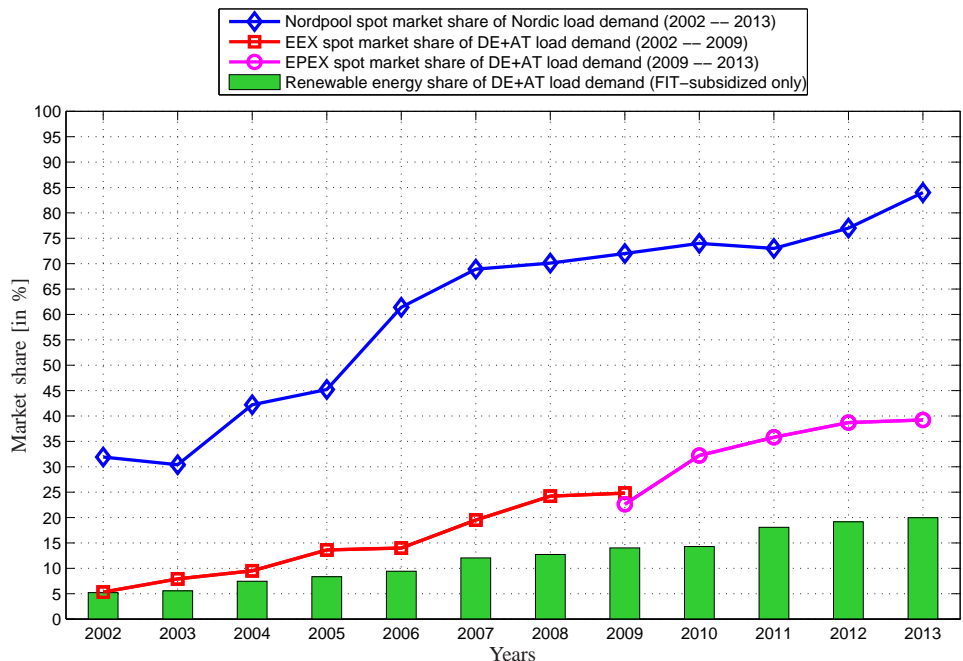}
    %\centerline{\psfig{figure=matlab/figures/modifiedAskCurve, width=1\linewidth} }
    \vspace{-0.20cm}
    \caption{Evolution of day-ahead market shares in the EEX/EPEX~German/Austrian market zone compared to the \emph{Nord Pool} spot market.}
    \label{fig:spotvolume_comparison}
    \vspace{-0.30cm}
\end{figure}

The stark contrast between the two power markets is that only around two-fifth ($\approx 39\%$ in~2013) of the yearly load demand in the German/Austrian grid zone is traded on the in \ac{epex}~day-ahead market, more than four-fifth ($\approx 85\%$ in~2013) of the load demand of the \emph{Nord Pool} countries (except Estonia) is traded on the day-ahead market.\newline 
The evolution of the EPEX day-ahead (EEX~spot until~2009) and \emph{Nord Pool} day-ahead market shares of the yearly load demand in their respective market zones is illustrated in \figurename~\ref{fig:spotvolume_comparison}. 
\begin{figure*}[t]
    %\vspace{-0.20cm}
    \centering
	\includegraphics[height=5.75cm,draft=false]{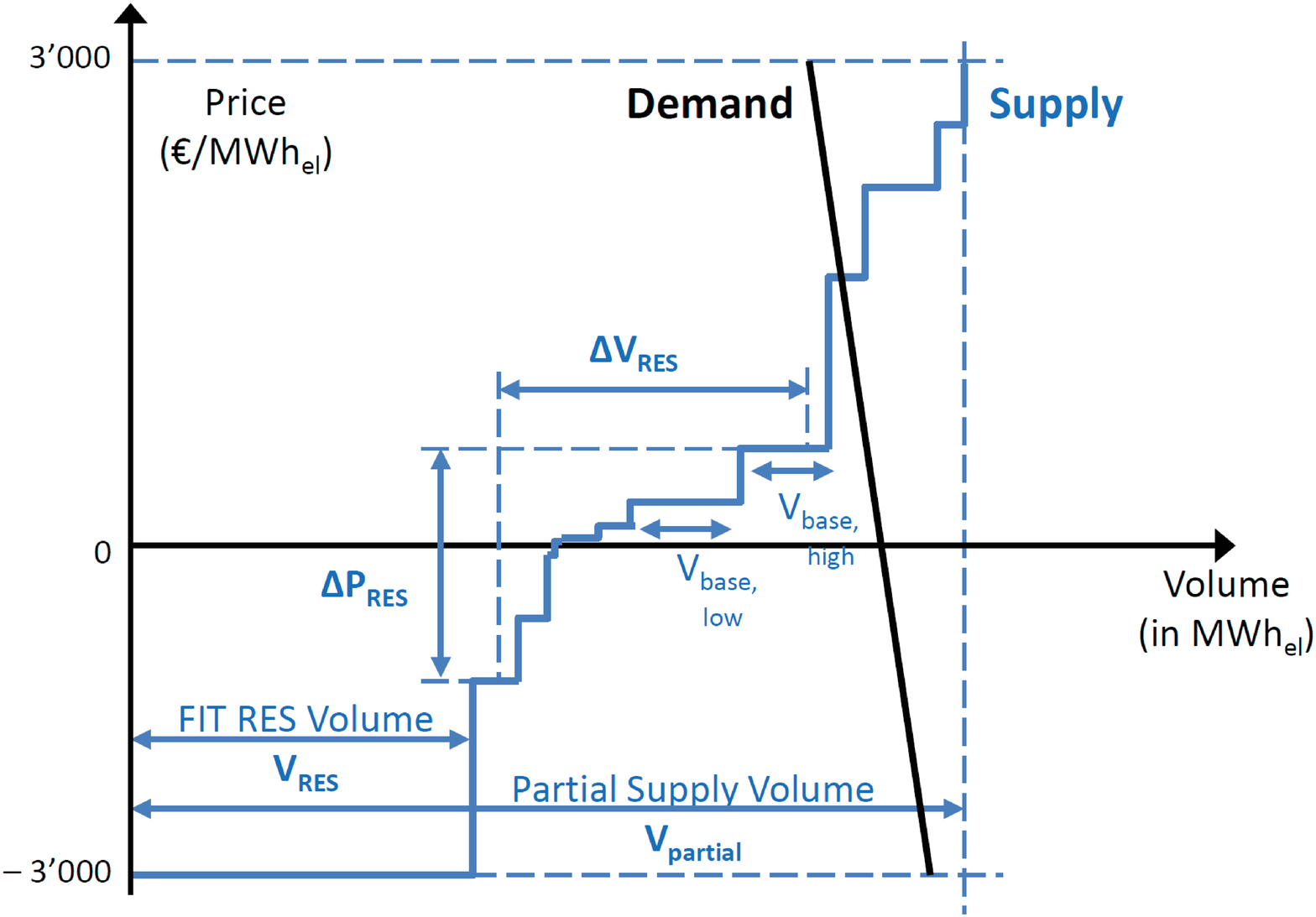}
	\includegraphics[height=5.80cm,draft=false]{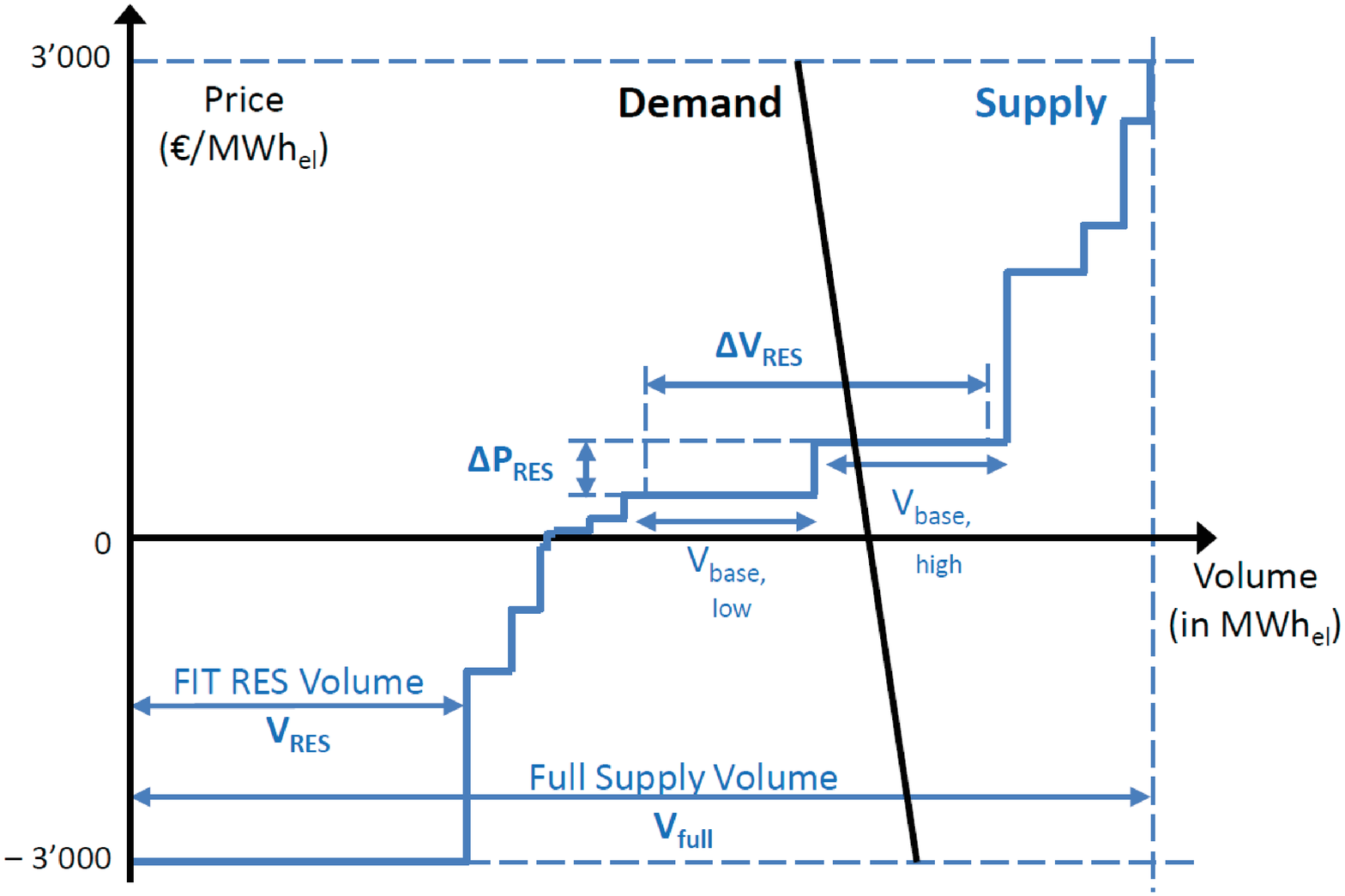}
    %\centerline{\psfig{figure=matlab/figures/modifiedAskCurve, width=1\linewidth} }
    \vspace{-0.30cm}
    \caption{Qualitative structure of merit-order bid/ask curves: (a)~in the current \ac{epex} day-ahead spot market, (b)~in a mandatory pool.}
    \label{fig:merit_order_change}
    \vspace{-0.3cm}
\end{figure*}
Market shares of both day-ahead markets have significantly grown over the years, but the \emph{Nord Pool} day-ahead market has a sizable lead in market share over \ac{epex}. Please note that not necessarily all of the FIT-subsidized RES feed-in in Germany and Austria ($\approx 20\%$ in~2013) may be given directly to the \ac{epex} day-ahead market,~i.e., as supply-side bids. Since~2010 there is an obligation on German \ac{tsos} to put all FIT-/EEG-subsidized electricity directly onto the day-ahead market. However, due to the increase of so-called \emph{direct marketing} of \ac{res} feed-in by private actors, notably wind power, the data situation is not entirely transparent. Nevertheless, it is fair to assume that all variable \ac{res} feed-in from wind \& \ac{pv} units -- by it's very nature a short term commodity -- influences the day-ahead spot market at least indirectly,~i.e., in the form of reduced load demand.
To our best knowledge, there is no inherent limitation and, hence, credible reason why similar spot market shares as in the Nordic countries could not be reached also in Germany, Austria or elsewhere in Europe.\newline 
One policy instrument for increasing the market share of the day-ahead market would be the introduction of a mandatory pool for physical energy procurement as is the case in Spain~(OMEL) and the US~power markets (e.g.~PJM)~\cite[p.~71, p.~106]{MarketBook:2008}, at least during a transitional phase, and the use of the futures market for hedging.

By doubling the \ac{epex} day-ahead market share, the \ac{fit}-subsidized \ac{res} feed-in share at the day-ahead market is, on average, cut in half. Quite obviously this would lead to a significant reduction of the well-known merit-order effect of \ac{res} power production. 
Structurally, the reduction of the merit-order effect is due to the on average lower slope $\left( \frac{d P_{\textrm{spot}}}{d V_{\textrm{spot}}} \right)$ of the supply bid curve. Due to the larger overall supply volume ($V_{\textrm{full}} > V_{\textrm{partial}}$) especially regarding base load generation units ($V_{\textrm{base, low}},\ V_{\textrm{base, high}}$), the spot price volatility $\Delta P_{\textrm{RES}}$ due to the volatility of the \ac{res} production volume $\Delta V_{\textrm{RES}}$ is significantly lower as is illustrated in Fig.~\ref{fig:merit_order_change}.

With the following assumptions:
\vspace{-0.10cm}
\begin{enumerate}
	\item[\textbf{I}] the full spot price range, i.e.~[$\pm$3'000\euro/MWh$_{\textrm{el}}$], is already covered in today's spot market supply curve (confer to Fig.~\ref{fig:modifiedBidAskCurve} for an actual bid/ask curve example),
	\item[\textbf{II}] under a mandatory pool regime, spot market supply volume will be higher than today, i.e.~$V_{\textrm{full}} > V_{\textrm{partial}}$, 
\end{enumerate}
the supply curve slope in a mandatory spot market pool has to be, on average, lower than today (or at most equal):
\begin{equation}
	\left( \frac{d P_{\textrm{spot}}}{d V_{\textrm{spot}}} \right)_{V_{\textrm{full}}} \le \left( \frac{d P_{\textrm{spot}}}{d V_{\textrm{spot}}} \right)_{V_{\textrm{partial}}} \quad.
\end{equation}
\vspace{-0.10cm}
The quantitative effect of the \ac{res} feed-in share at the day-ahead market on the merit-order effect is part of subsequent investigations in~Section~\ref{sec:EffectMarginalCosts}.

\section{Effects of including RES Marginal Costs in Power Market Ask Curve}\label{sec:EffectMarginalCosts}
%In \ref{sec:QuantAnalysis} the impact of \ac{res} with \ac{fit} on the current market and in comparison to the overall volume is discussed, this section discusses the effect on the day-ahead prices under the assumption, that wind and \ac{pv} feed-in has to be settled at the day-ahead market with marginal costs, but the volume risk of the forecast error (predicted feed-in versus the realized feed-in) has not to be covered by the \ac{res} supplier (this is still a support scheme, but a less strong than \ac{fit}).\\

In the following we analyze the quantitative effect on day-ahead prices if wind and \ac{pv} feed-in would have to be settled at the day-ahead market using the marginal operation costs of \ac{res} units. The volume risk of the forecast error,~i.e., predicted feed-in versus realized feed-in, is not covered by the \ac{res} supplier. Note, that this situation would still constitute a support scheme but a milder one than the existing \ac{fit} scheme.
Compared to a fossil-fueled power plant, ``fuel´´ costs of wind \& \ac{pv} plants or any other variable \ac{res} production units such as run-of-river hydro plants or the solar field of \ac{csp} plants are obviously zero. (Note that a gas-fueled part of a \ac{csp} plant would obviously incur fuel costs.)

There are, however, other cost components that can be considered as marginal operation costs for variable \ac{res} units. These can include marginal wear \& tear from plant operation, at least the part that can be directly attributed to run-time and energy produced, for any turbine-driven plant as well as concession taxes and land-lease payments that are calculated per unit energy produced~(in $\textrm{MWh}_\textrm{el}$). Energy-based marginal concession taxes exist, for example, for hydro units in Vallais, Switzerland (\emph{Wasserzins}) and Saxony, Germany (\emph{Wasserentnahmeabgabe}). Land-lease payments,~e.g., for the land-use of wind \& \ac{pv} units are at the discretion of the involved parties,~i.e., land-owner and \ac{res} plant owner, but often do include a revenue component for the electricity plant.

Mid-term, e.g.,~by the time existing \ac{fit}-support schemes are phased-out or adapted to not include a grid feed-in guarantee, it is likely that at least a part of the day-ahead price and volume risk of RES feed-in has to be covered by these units, notably wind \& \ac{pv}, directly. Such an obligation would result in a form of an insurance fee or risk coverage for the forecast error of the variable \ac{res} electricity production. Today, system operators usually do have to cover the forecast error by acquiring control reserve power capacity \emph{a priori}, which then provides the needed balancing energy in real-time.\newline
Total marginal operation cost of running variable \ac{res} units is thus the sum of the individual cost components:%\newline
%\begin{center}
%	Marginal production costs of RES unit~[\euro/MWh$_\textrm{el}$]
%\end{center} 
%\vspace{-0.25cm}
\begin{IEEEeqnarray}{lcl}
%\vspace{-0.25cm}
 	\textrm{Marginal production costs of RES unit [\euro/MWh$_\textrm{el}$] }
 								   &=&\ \textrm{wear \& tear cost [\euro/MWh$_\textrm{el}$]} \nonumber \\
								   &+&\ \textrm{land-lease cost [\euro/MWh$_\textrm{el}$]} \nonumber \\
								   &+&\ \textrm{concession tax [\euro/MWh$_\textrm{el}$]}  \nonumber \\
								   &+&\ \textrm{forecast error [\euro/MWh$_\textrm{el}$]}  \ . \nonumber
\end{IEEEeqnarray}

The effect is that the marginal operational costs for \ac{res} production, especially wind \& \ac{pv} units, become positive, since at least the risk coverage costs must be earned by the actual electricity production. This, obviously, will change the bidding behavior of wind \& \ac{pv} units as well as the structure of the merit-order curve. For a system's perspective on RES grid integration costs, confer to~\cite{hirth2012}.

\subsection{Literature Review -- Marginal Costs for Wind\&PV units}\label{sec:marCostLiterature}
We obtained general wear\&tear operation costs of \ac{res} units, essentially operation and maintenance (O$\&$M) costs, from a literature survey of recent reports on electricity generation costs \cite{IEA2012aa, Eurelectric2012, FHGISE2013}. Operation cost for wind (onshore) generation were in the range of $15$--$27$~\euro/MWh$_\textrm{el}$, whereas for \ac{pv} generation costs were in the range of $22$--$33$~\euro/MWh$_\textrm{el}$. Within OECD countries median O$\&$M values of $17$~\euro/MWh$_\textrm{el}$ for onshore wind and of $23$~\euro/MWh$_\textrm{el}$ for \ac{pv} have been presented \cite[p.~102]{IEA2012aa}. The largest share of these operation costs will not be truly marginal,~i.e., in the sense that additional cost will be incurred for every additional (marginal) electricity unit (in MWh$_\textrm{el}$) produced. For the moving parts of wind turbines etc. one can fairly attribute some marginal O\&M cost, whereas on can probably not do so in the case of \ac{pv} units.

Marginal costs for land-lease as well as concession taxes will always be site-specific. For the above mentioned hydro concession taxes the cost range is about~$10$--$20$~\euro/MWh$_\textrm{el}$.

The marginal costs or insurance fee for covering the risk of forecast errors and the involved costs for calling on control reserve power and energy via ancillary service markets will also be site-specific and depend on the responsible system operator. In the case of the German \acp{tsos}, an estimation of balancing costs for wind \& \ac{pv} forecast errors is possible due to recent regulatory transparency efforts. Balancing costs have recently been assessed to be in the range of $0$--$10$~\euro/MWh$_\textrm{el}$ for the years 2011--2013~\cite{mueller:2014}.

\subsection{Wind \& PV Supply Bid Assumptions}\label{sec:marCostAssumptions}
The subsequent analysis is based on certain assumptions regarding wind \& \ac{pv} marginal costs and supply bids:
\begin{itemize}
\item In case the \ac{fit} scheme provides a feed-in guarantee for all \ac{fit}-supported \ac{res} production: we assume that the \ac{tso} gives all \ac{res} with a large negative price bid,~i.e., --3000~\euro/MWh$_\textrm{el}$ to the day-ahead market, to ensure the settlement.
\item In case the \ac{fit} scheme does not provide a feed-in guarantee for all \ac{fit}-supported \ac{res} production: we assume total marginal operation costs for both wind \& \ac{pv} units to be in the range of 0--25~\euro/MWh$_\textrm{el}$ and simulate in steps of 5~\euro/MWh$_\textrm{el}$.
\item Only wind and \ac{pv} power feed-in are considered since they represent the majority,~i.e., two-third, of German \ac{fit}-subsidized \ac{res} feed-in.
\item There is no separation between on-shore and off-shore wind power feed-in. Currently less than 5\% of the installed wind capacity in Germany is off-shore and therefore negligible at the moment. In the future two different O\&M costs for off-shore and on-shore wind power feed-in have to be considered.
\item The demand side of power markets is not affected.
\end{itemize}

%\textbf{To-Dos}
%\begin{enumerate}
%	\item Grid Integration Costs
%        \begin{itemize}
%        \item Sensitivity range for marginal costs for Wind \& \ac{pv}: 0-20~€/MWh in steps of 5~€/MWh.
%        \item Average day-ahead market price as a function of merit-order bid for wind \& \ac{pv}.
%        \item Grid Integration: Annual share of integrated RES energy (e.g.~95\%) as a function of merit-order bid for wind \& \ac{pv}. (Outlook for later)
%        \end{itemize}
%
%    \item Good arguments why Wind \& \ac{pv} actually do have marginal costs
%        \begin{itemize}
%        \item Provide equation with possible cost sources, grid integration costs [control reserve power procurement, power provision, variable Pachtzahlungen in Funktionen des Umsatzes], Wasserzins/abgabe [based on actual energy production, CH-Vallais, DE-Saxonie])
%        \item Reference to costs for procuring balancing power \& energy in DE over last years [0-10~€/MWh](MA Jonas Müller)
%        \end{itemize}
%
%    \item Statistics Update (Year 2013, December Work, Estimates when possible, EPEX volumes versus Nordpool volumes)
%
%    \item Discussion of whether all RES feed-in actually turned up at day-ahead market directly/indirectly (short review of merit-order effect literature, underlying assumption is that all FIT-Infeed has and indirect influence on day-ahead market clearing, if not much of the relevant literature would give false impressions, process not very transparent, therefore clear statements are difficult)

\subsection{Method}
Wind/\ac{pv} \ac{fit} feed-in is settled over the day-ahead market with an ask of --3000~\euro/MWh$_\textrm{el}$, since the \ac{fit} paid is independent from the achieved day-ahead spot market price and the \ac{tso} has to ensure the settlement of the \ac{res} feed-in. One of the main questions is now, what would be the resulting market prices, if wind \& \ac{pv} would have to be settled under market conditions. In the following we investigate the impacts of settling \ac{res} feed-in with realistic marginal costs, as given in Section \ref{sec:marCostLiterature}.

\begin{figure}[tb]
\centering
    %\centerline{\psfig{figure=figures/ComparisonPeakShapeYearly.eps,width=\linewidth} }
    \psfrag{x}[cc][cc]{\scriptsize\shortstack{Volume [MWh$_\textrm{el}$]}}
    \psfrag{y}[cc][cc]{\scriptsize\shortstack{Price [\euro/MWh$_\textrm{el}$]}}
    \includegraphics[width=0.60\textwidth]{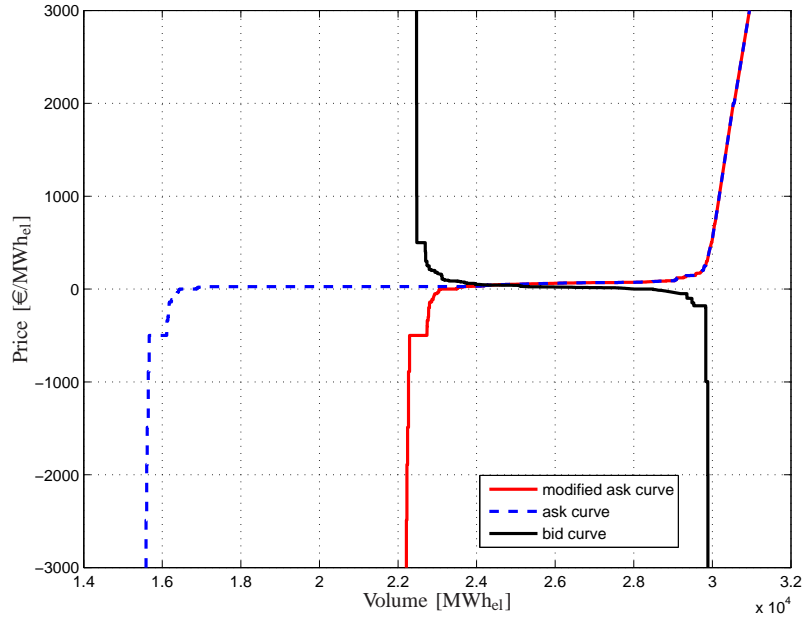}
    %\centerline{\psfig{figure=matlab/figures/modifiedAskCurve, width=1\linewidth} }
    \caption{Modified \ac{epex} spot market bid/ask curve (RES~bids~=~$25$\euro/MWh$_\textrm{el}$), 1~January~2011 (1$^{\textrm{st}}$~hour).}
    \label{fig:modifiedBidAskCurve}
   %  \vspace{-0.5cm} % Journal version
        %\vspace{-0.25cm} % ArXiv version
\end{figure}

For the calculations we settle the day-ahead market based on the \ac{epex} bid-ask-curve. We leave the demand curve untouched and rearrange the ask curve by applying the corresponding marginal costs on wind \& \ac{pv} volume. The settlement is done based on the \ac{epex} settlement \cite{EuropeanEnergyExchange2012}. \figurename~\ref{fig:modifiedBidAskCurve} shows the volume of the realized and the modified ask curve. Here, settling wind \& \ac{pv} with assumed marginal costs results in a visible jump for wind energy bids and the resulting power market ask curve.
%
%\begin{figure}[h]
%\centering
%    %\centerline{\psfig{figure=figures/ComparisonPeakShapeYearly.eps,width=\linewidth} }
%    \psfrag{x}[cc][cc]{\scriptsize\shortstack{Bids}}
%    \psfrag{y}[cc][cc]{\scriptsize\shortstack{Bid volume [in MWh]}}
%    \includegraphics[width=0.45\textwidth]{modifiedAskCurve.eps}
%    %\centerline{\psfig{figure=matlab/figures/modifiedAskCurve, width=1\linewidth} }
%    \caption{Modified ask curve for 1$^{\textrm{st}}$ of January 2011 (hour 1).}
%    \label{fig:modifiedVolumeCurve}
%    \vspace{-0.3cm}
%\end{figure}
%

\subsection{Results}
For the analysis, we calculate artificial day-ahead prices assuming significant marginal costs,~i.e., $25\,$\euro/$\textrm{MWh}_{\textrm{el.}}$ of \ac{res} feed-in for the following scenarios:
\begin{enumerate}
\item Only wind feed-in is settled with marginal costs.
\item Only \ac{pv} feed-in is settled with marginal costs.
\item Wind \& \ac{pv} feed-in are settled with marginal costs.
\end{enumerate}
We discuss the effects that the resulting artificial day-ahead prices have on the three principal power plant types,~i.e., base load, flexible peak load plants and storage units.

\subsubsection{Analysis of Spot Price Formation}
\begin{figure}[p]
\centering
    %\centerline{\psfig{figure=figures/ComparisonPeakShapeYearly.eps,width=\linewidth} }
    \psfrag{x}[cc][cc]{\scriptsize\shortstack{Time [with hourly granularity]}}
    \psfrag{y}[cc][cc]{\scriptsize\shortstack{Day-ahead price [in \euro/MWh$_\textrm{el}$]}}
    \includegraphics[width=0.675\textwidth]{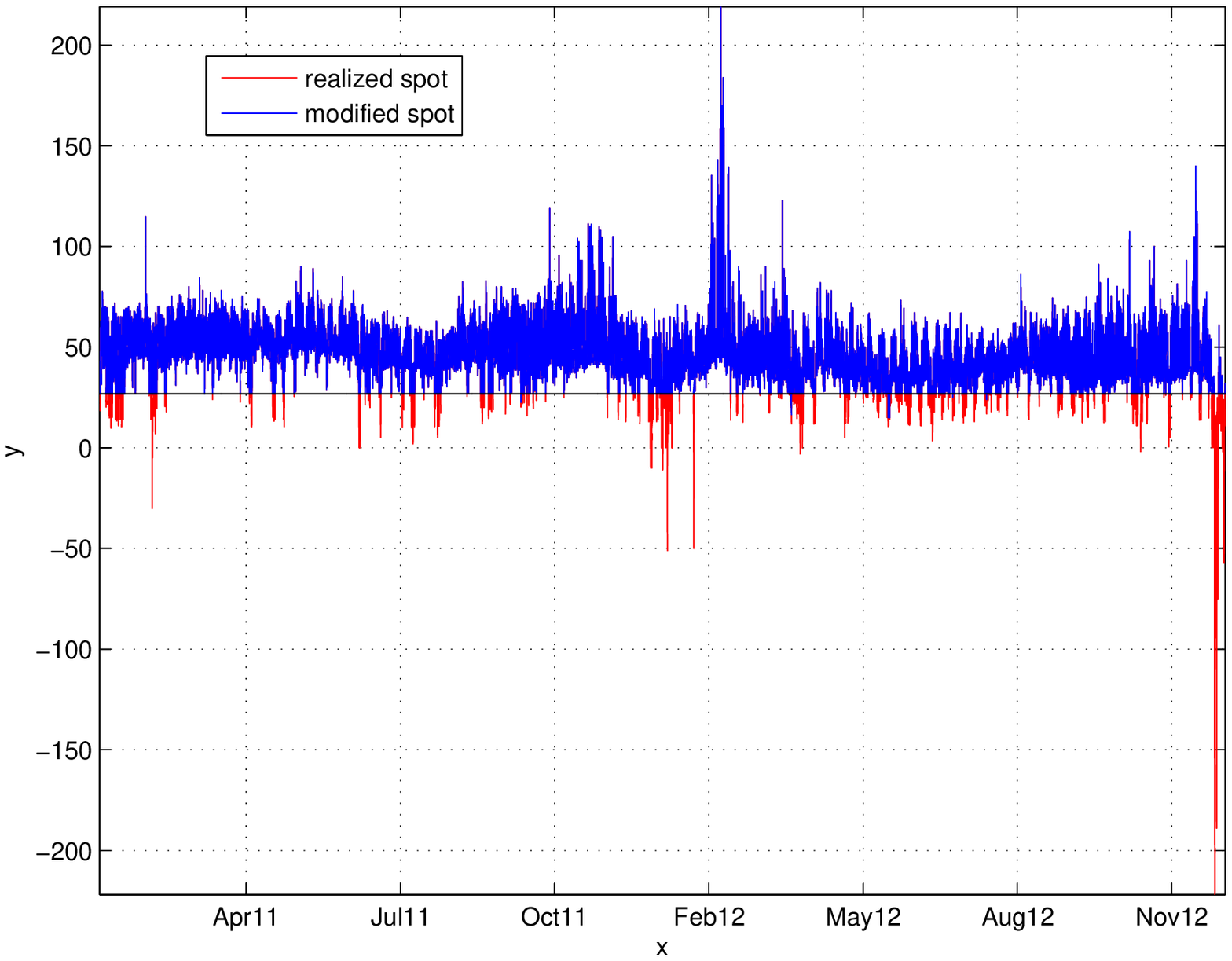}
    %\centerline{\psfig{figure=matlab/figures/modifiedAskCurve, width=1\linewidth} }
    \caption{Comparison of realized day-ahead spot prices without (\ac{fit}) and with marginal operation costs ($25$\euro/MWh$_\textrm{el}$); wind feed-in only.}
    \label{fig:newPriceWithWind}
%        \vspace{-0.5cm} % Journal version
      %  \vspace{-0.25cm} % ArXiv version
\end{figure}

First, we analyze the effect when wind feed-in is settled with marginal costs whereas \ac{pv} feed-in is still subsidized. \figurename~\ref{fig:newPriceWithWind} shows the settlement of day-ahead prices under the assumption, that wind energy is settled with marginal costs. It is shown, that no negative price effects are existing anymore. Some hours are settled below the marginal cost of wind units. In these hours the power feed-in from other, cheaper sources is sufficient for covering the load demand; no wind power was dispatched.

\begin{figure}[p]
\centering
    %\centerline{\psfig{figure=figures/ComparisonPeakShapeYearly.eps,width=\linewidth} }
    \psfrag{x}[cc][cc]{\scriptsize\shortstack{Time [with hourly granularity]}}
    \psfrag{y}[cc][cc]{\scriptsize\shortstack{Day-ahead price [in \euro/MWh$_\textrm{el}$]}}
    \includegraphics[width=0.675\textwidth]{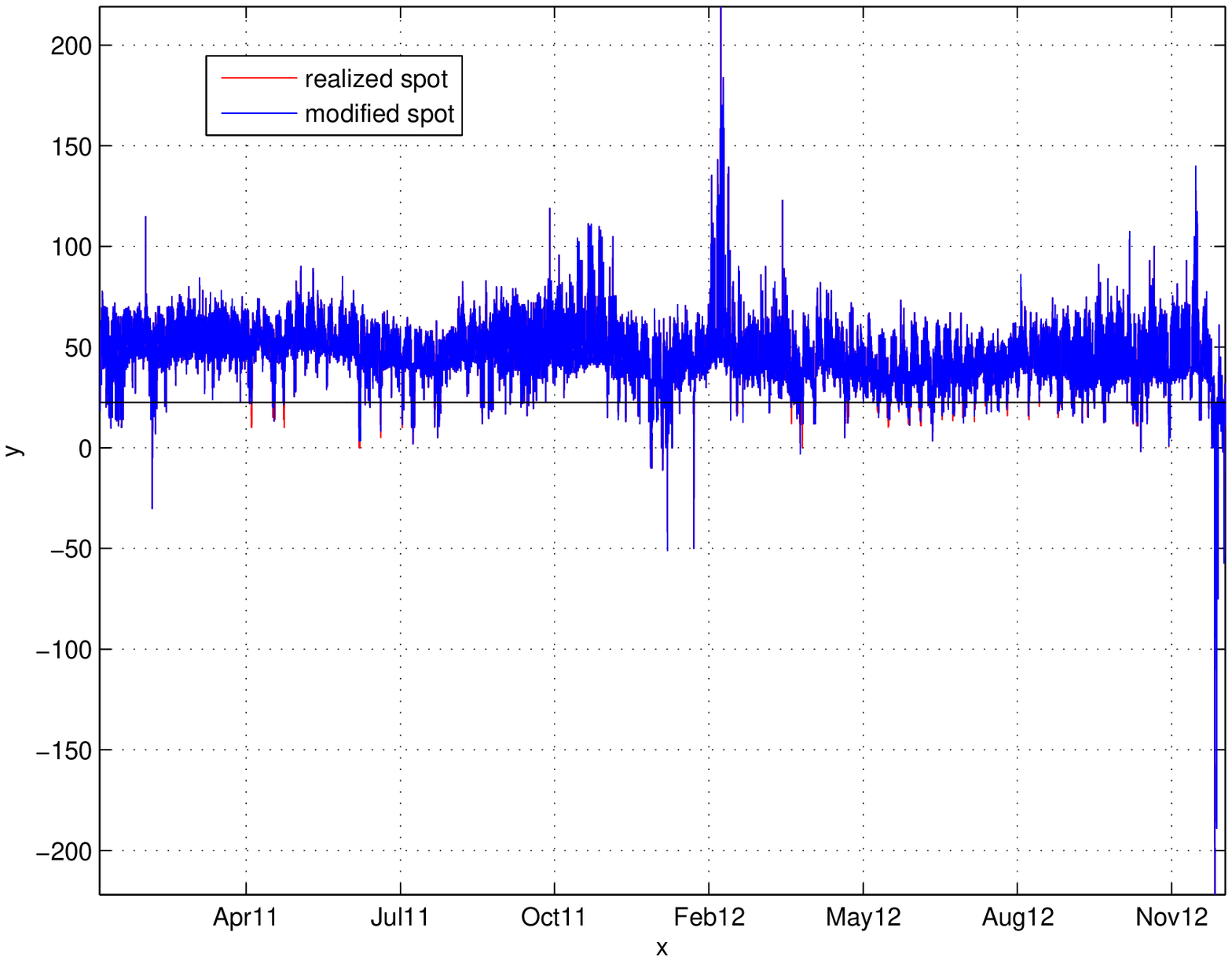}
    %\centerline{\psfig{figure=matlab/figures/modifiedAskCurve, width=1\linewidth} }
    \caption{Comparison of realized day-ahead with \ac{fit} and day-ahead with marginal costs (RES bids=$25$\euro/MWh$_\textrm{el}$); \ac{pv} feed-in only.}
    \label{fig:newPriceWithPV}
    %\vspace{-0.5cm} % Journal version
     %   \vspace{-0.25cm} % ArXiv version
\end{figure}

\figurename~\ref{fig:newPriceWithPV} shows the constructed day-ahead prices where wind feed-in is still supported with \ac{fit} and \ac{pv} is settled using marginal costs. The analysis shows, that the pure \ac{pv} feed-in over marginal costs does not change the overall settlement of high prices above the marginal cost level. The spikes in the day-ahead prices remain and so do negative day-ahead price events. The negative spot prices do not vanish in the pure \ac{pv} scenario because they are all occurring during off-peak times, where \ac{pv} production coincides with high demand. If the installed \ac{pv} capacity increases further, we may observe negative prices during peak hours as a result of \ac{pv} feed-in as well.

\begin{figure}[tbh]
\centering
    %\centerline{\psfig{figure=figures/ComparisonPeakShapeYearly.eps,width=\linewidth} }
    \psfrag{x}[cc][cc]{\scriptsize\shortstack{Time [with hourly granularity]}}
    \psfrag{y}[cc][cc]{\scriptsize\shortstack{Day-ahead price [in \euro/MWh$_\textrm{el}$]}}
    \includegraphics[width=0.675\textwidth]{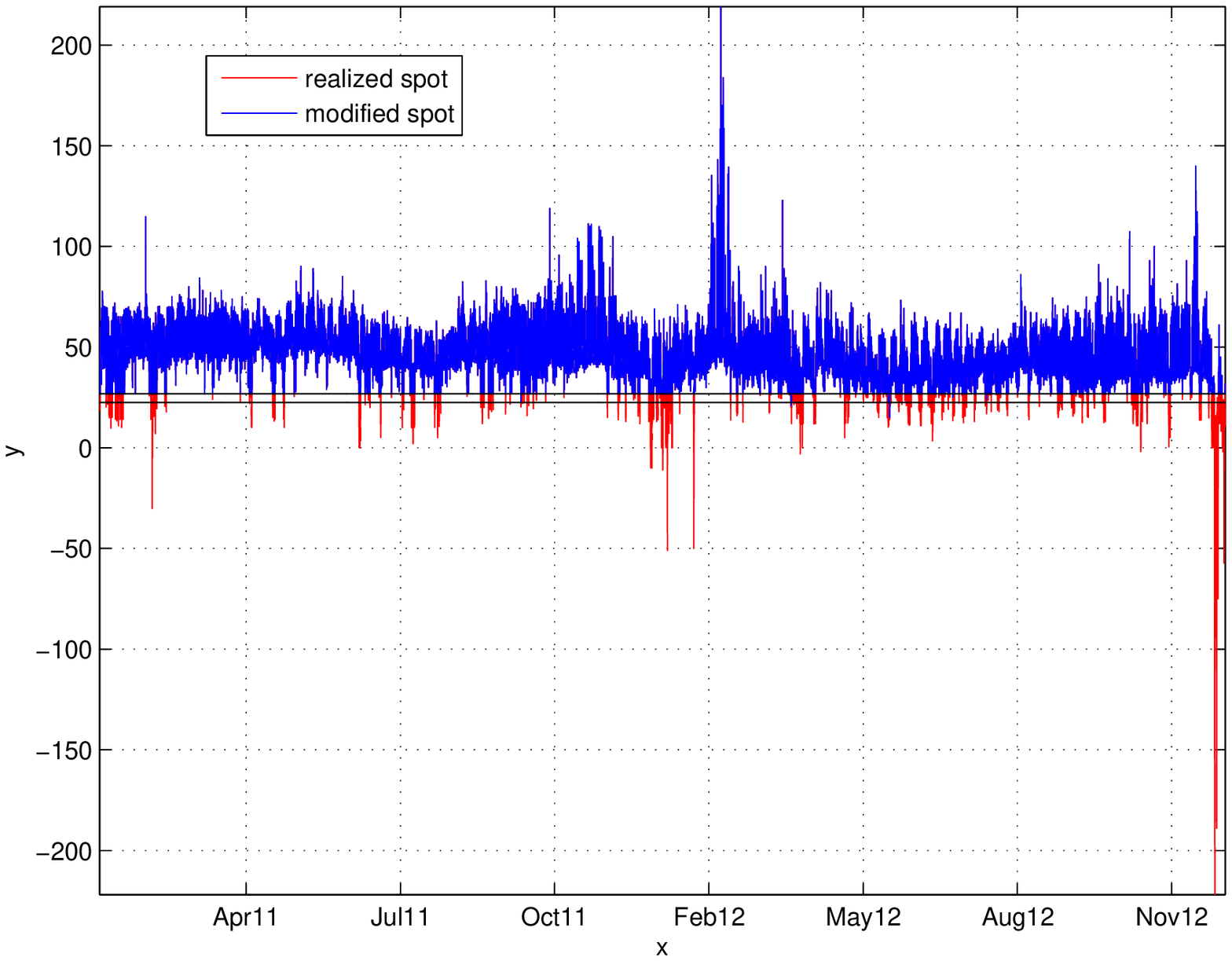}
    %\centerline{\psfig{figure=matlab/figures/modifiedAskCurve, width=1\linewidth} }
    \caption{Comparison of realized day-ahead with \ac{fit} and day-ahead with marginal costs (RES bids=$25$\euro/MWh$_\textrm{el}$) (wind \& \ac{pv} feed-in).}
    \label{fig:newPriceWithWindAndPV}
%        \vspace{-0.5cm} % Journal version
        %\vspace{-0.25cm} % ArXiv version
\end{figure}

\figurename~\ref{fig:newPriceWithWindAndPV} shows the bid/ask curves using marginal costs for wind \& \ac{pv} power feed-in. As for wind feed-in only, some hours exist below the marginal costs of \ac{pv}, which are lower than the wind marginal costs. In those hours, production from other base load power plants is sufficient; no \ac{res} power was dispatched.\\

\subsubsection{Producer Perspective}
As discussed in Section \ref{sec:introduction}, we analyze the effect of the settlement of wind \& \ac{pv} feed-in on three groups of power plants:
\begin{enumerate}
\item Base load power plants (nuclear, lignite, hard coal),
\item Peak load power plants (gas and storage plants),
\item Short term storage (\ac{pshp} and other technologies).
\end{enumerate}

As shown previously in \figurename~\ref{fig:newPriceWithWindAndPV}, all negative prices vanish as soon as \ac{res} units have to settle with their respective marginal costs. In some hours the settled day-ahead price is even lower than the marginal cost of \ac{pv} and covered by other base load plants. In addition, the marginal costs of at least some base load power plants are below the marginal cost of wind \& \ac{pv} units. 
The existence of bids below the introduced marginal costs for wind \& PV indicates that conventional base load power plants have marginal costs below the marginal costs for wind \& PV. Besides negative marginal costs, which is often the result of avoiding otherwise necessary power plant down-ramping or even shutdowns in some hours, also longer periods of no wind \& PV production exist and therefore large base load power plants can run profitable below the in this paper assumed marginal costs for wind \& PV.\\
The second category of plants, flexible fossil-fueled plants like natural gas fired units and larger storage plants, are not affected by the change, since they profit from peak hours.

%\figurename \ref{fig:newPriceWithDoublePV} shows a simulation with a linear scaling of the realized \ac{pv} feed-in, where the feed-in is doubled in every hour. The peak prices are not affected by this manipulation, neither are negative price events which still prevail.
%%
%\begin{figure}[tb]
%\centering
%    %\centerline{\psfig{figure=figures/ComparisonPeakShapeYearly.eps,width=\linewidth} }
%    \psfrag{x}[cc][cc]{\scriptsize\shortstack{Time [with hourly granularity]}}
%    \psfrag{y}[cc][cc]{\scriptsize\shortstack{Spot price [in \euro/MWh]}}
%    \includegraphics[width=0.45\textwidth]{newPriceWithDoublePV.eps}
%    %\centerline{\psfig{figure=matlab/figures/modifiedAskCurve, width=1\linewidth} }
%    \caption{Comparison of realized day-ahead with \ac{fit} and and day-ahead with marginal costs (twice the current \ac{pv} feed-in).}
%    \label{fig:newPriceWithDoublePV}
%    \vspace{-0.3cm}
%\end{figure}
%
%The short term energy storage facilities are most affected. Those plants generate profit by statistical electricity price arbitrage; storing energy during low price hours and producing during high price hours. Since high prices are not affected but lower prices are increasing, hence the day-ahead price spread is decreasing, the business model is even more under pressure than it is already today \cite{Hildmann2011a}.

%\begin{figure}[b] % Journal version
\begin{figure}[p] % ArXiv version
	\centering
    \psfrag{x}[cc][cc]{\scriptsize\shortstack{Time [with hourly granularity]}}
    \psfrag{y}[cc][cc]{\scriptsize\shortstack{Day-ahead price [in \euro/MWh$_\textrm{el}$]}}
    \includegraphics[width=0.675\textwidth]{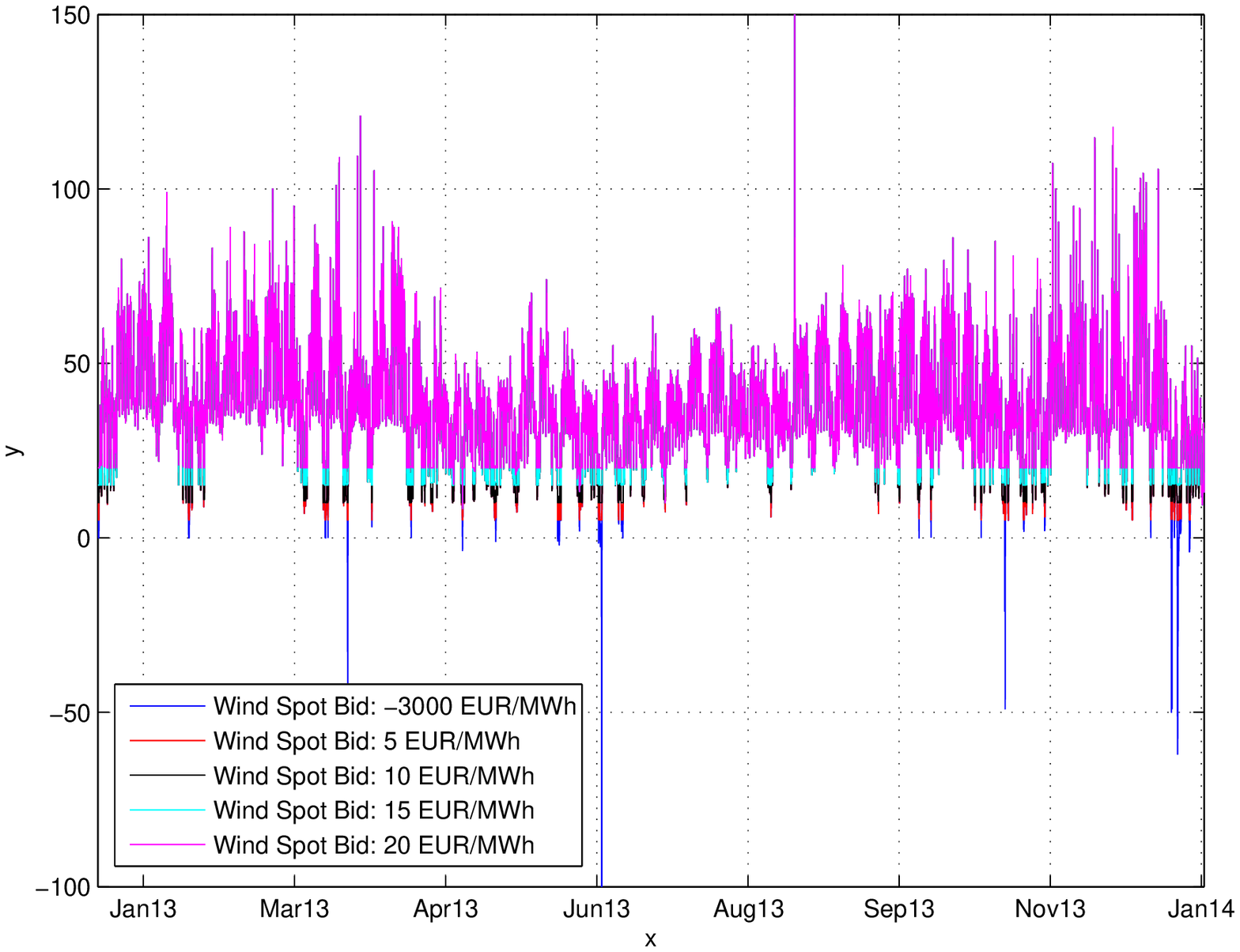}
    \psfrag{x}[cc][cc]{\scriptsize\shortstack{Time [with hourly granularity]}}
    \psfrag{y}[cc][cc]{\scriptsize\shortstack{Day-ahead price [in \euro/MWh$_\textrm{el}$]}}
    \includegraphics[width=0.675\textwidth]{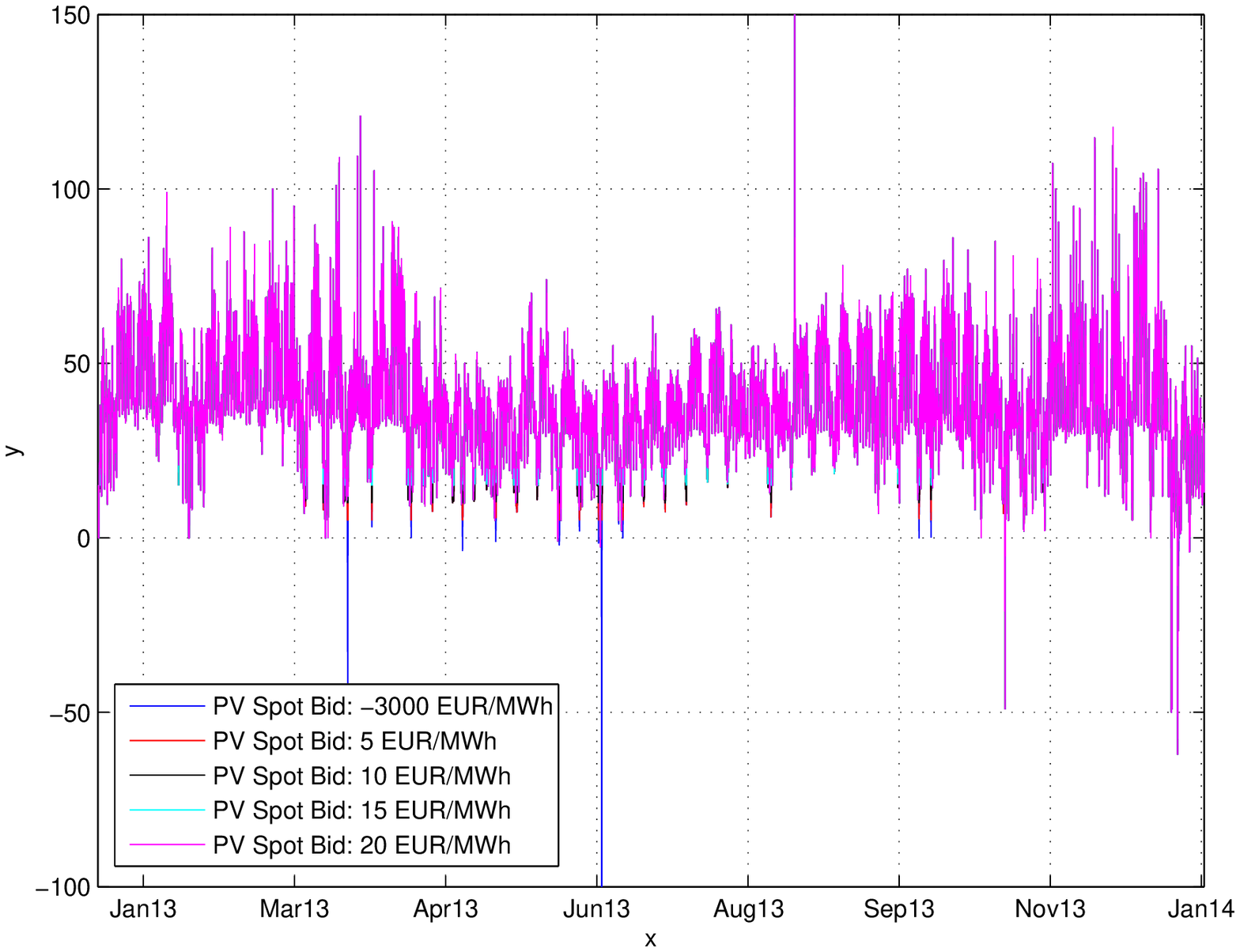}
    \caption{Comparison of realized day-ahead with \ac{fit} and with varying marginal costs (0--20\euro/MWh$_\textrm{el}$): (a)~only wind, (b)~only \ac{pv}.}
    \label{fig:MarginalCostRangePVWind}
%        \vspace{-0.5cm} % Journal version
       % \vspace{-0.25cm} % ArXiv version
\end{figure}

\subsubsection{Consumer Perspective}
The high costs generated by the \ac{fit} will be lowered, because the \ac{res} generation can be settled at a significantly higher price. Costs for the coverage of the forecast error of wind and \ac{pv} generators are still present, but determining the incurred costs due to this is clearly beyond the scope of the paper. Given the long-term guarantee of the \ac{fit} scheme (20 years), the overall \ac{res} deployment cost can be lowered significantly and a transition to a competitive spot market seems possible.

\subsubsection{RES Producer Perspective under Market Constraints}
As shown in \figurename~\ref{fig:modifiedBidAskCurve} and \ref{fig:newPriceWithWindAndPV}, a large amount of \ac{res} feed-in is still settled via the day-ahead spot market, if \ac{res} feed-in is given to the market with marginal costs. While during the \ac{fit} period, only the costs for \ac{fit} will be lowered, a real market situation, independent from the prediction error, seems also possible. Future work will address the impact of such a regime change, which gives no \ac{res} feed-in guarantee, for the effective \ac{res} grid integration.
%
%\begin{figure}[h]
%    %\centerline{\psfig{figure=figures/ComparisonPeakShapeYearly.eps,width=\linewidth} }
%    \psfrag{x}[cc][cc]{\scriptsize\shortstack{\euro/MWj}}
%    \psfrag{y}[cc][cc]{\scriptsize\shortstack{MW}}
%    \includegraphics[width=0.45\textwidth]{matlab/figures/newPriceWithWindDoublePV.eps}
%    %\centerline{\psfig{figure=matlab/figures/modifiedAskCurve, width=1\linewidth} }
%    \caption{Comparison of realized day-ahead with \ac{fit} and and day-ahead with marginal costs with wind double the current \ac{pv} feed-in}
%    \label{fig:meanHoursYear}
%    \vspace{-0.3cm}
%\end{figure}

%\begin{figure}[h]
%    %\centerline{\psfig{figure=figures/ComparisonPeakShapeYearly.eps,width=\linewidth} }
%    \psfrag{x}[cc][cc]{\scriptsize\shortstack{\euro/MWj}}
%    \psfrag{y}[cc][cc]{\scriptsize\shortstack{MW}}
%    \includegraphics[width=0.45\textwidth]{matlab/figures/newPriceWithWindNPV.eps}
%    %\centerline{\psfig{figure=matlab/figures/modifiedAskCurve, width=1\linewidth} }
%    \caption{Comparison of realized day-ahead with \ac{fit} and and day-ahead with marginal costs with wind double the current \ac{pv} feed-in}
%    \label{fig:meanHoursYear}
%    \vspace{-0.3cm}
%\end{figure}
%\end{enumerate}

\subsection{Sensitivity Analysis of Marginal Operation Costs}
Since the real marginal costs of wind \& \ac{pv} are site-specific and thus hard to estimate, we have compiled a sensitivity analysis of the impacts of different assumed marginal operation costs, ranging from 0--20~\euro/MWh$_\textrm{el}$ in steps of 5~\euro/MWh$_\textrm{el}$, for either wind or \ac{pv} power feed-in.

The effects of these assumed marginal costs on the day-ahead market clearing are illustrated in \figurename~\ref{fig:MarginalCostRangePVWind}. A quantitative analysis of average day-ahead spot prices and their volatility is given in Tab.~\ref{tab:meanPV}--\ref{tab:stdWind}. While the effects increase with higher marginal costs, the overall observations of the paper are nevertheless validated.
As can be seen, average day-ahead spot market prices will be higher if wind or \ac{pv} feed-in is bid into the day-ahead market with positive marginal costs. Also, the price volatility (i.e., ~standard deviation) is decreasing. Both effects are significantly more pronounced in the case of wind power feed-in, as shown in~\figurename~\ref{fig:MarginalCostRangePVWind}~(a) clearly shows that positive marginal costs for wind feed-in push the day-ahead clearing prices above zero in all instances. This is not the case for \ac{pv} feed-in, see ~\figurename~\ref{fig:MarginalCostRangePVWind}~(b). The original clearing results for an feed-in guarantee via highly negative bids (--3000~\euro/MWh$_\textrm{el}$) are also shown.
\begin{table}[t]
\begin{center}
	\renewcommand{\arraystretch}{1.0}
	\centering
	
	\footnotesize
    \caption{Average day-ahead spot price}% \newline (marginal costs for \ac{pv} feed-in only)}
    \label{tab:meanPV}
    %\vspace{-0.30cm} % ArXiv version
    \begin{tabular}{l|ccc}
      \textbf{Marginal cost of PV}                       &     \textbf{Year} \\
      \textbf{(\euro/MWh$_\textrm{el}$)} 								&     \textbf{2011}    &   \textbf{2012}  &  \textbf{2013} \\
     \hline
-3000   & 51.5 &   42.8  & 37.9            \\
     5         & 51.55 & 42.8  &  38.0  \\
     10       &    51.6 & 42.8 & 38.0   \\
     15      &       51.6 & 42.8 & 38.1\\
     20      &      51.6 & 42.9 & 38.3 \\
\hline
    \end{tabular}
    %\end{small}
%        \vspace{-0.5cm} % Journal version
        \vspace{-0.30cm} % ArXiv version
\end{center}
\end{table}
\begin{table}[!h]
\begin{center}
	\renewcommand{\arraystretch}{1.0}
	\centering
	
	\footnotesize
    \caption{Standard deviation of day-ahead spot price}% \newline (marginal costs for \ac{pv} feed-in only)}
    \label{tab:stdPV}
    %\vspace{-0.30cm} % ArXiv version
    \begin{tabular}{l|ccc}
     \textbf{Marginal cost of PV}                       &     \textbf{Year} \\
      \textbf{(\euro/MWh$_\textrm{el}$)} 								&     \textbf{2011}    &   \textbf{2012}  &  \textbf{2013} \\
     \hline
-3000   & 13.4 &  18.6  & 16.5           \\
     5         & 13.4 & 18.6  &  16.2  \\
     10       &    13.3 & 18.6 &16.2   \\
     15      &       13.3 & 18.6 & 16.0\\
     20      &      13.3 & 18.5 & 15.8 \\
\hline
    \end{tabular}
    %\end{small}
%        \vspace{-0.5cm} % Journal version
        \vspace{-0.30cm} % ArXiv version
\end{center}
\end{table}
\begin{table}[!h]
\begin{center}
	\renewcommand{\arraystretch}{1.0}
	\centering
	
	\footnotesize
	\caption{Average day-ahead spot price}% \newline (marginal costs for wind feed-in only)}
    \label{tab:meanWind}
    %\vspace{-0.30cm} % ArXiv version
    \begin{tabular}{l|ccc}
     \textbf{Marginal cost of Wind}                       &     \textbf{Year} \\
      \textbf{(\euro/MWh$_\textrm{el}$)} 								&     \textbf{2011}    &   \textbf{2012}  &  \textbf{2013} \\
     \hline
-3000   & 51.5 &   42.8  & 37.9            \\
     5         & 51.6 & 43.2  &  38.0  \\
     10       &    51.6 & 43.3 & 38.1   \\
     15      &       51.7 & 43.4 & 38.5\\
     20      &      51.8 & 43.6 & 38.9 \\
\hline
    \end{tabular}
    %\end{small}
%        \vspace{-0.5cm} % Journal version
        \vspace{-0.30cm} % ArXiv version
\end{center}
\end{table}
\begin{table}[!h]
\begin{center}
	\renewcommand{\arraystretch}{1.0}
	\centering
	
	\footnotesize
	\caption{Standard Deviation of day-ahead spot price}% \newline (marginal costs for wind feed-in only)}
    \label{tab:stdWind}
    %\vspace{-0.30cm} % ArXiv version
    \begin{tabular}{l|ccc}
     \textbf{Marginal cost of Wind}                       &     \textbf{Year} \\
      \textbf{(\euro/MWh$_\textrm{el}$)} 								&     \textbf{2011}    &   \textbf{2012}  &  \textbf{2013} \\
     \hline
-3000   & 13.4 &  18.6  & 16.5           \\
     5         & 13.3 & 15.8  &  15.9  \\
     10       &    13.2 & 15.7 &15.7   \\
     15      &       13.0 & 15.5 & 15.2\\
     20      &      12.7 & 15.0 & 14.5\\
\hline
    \end{tabular}
    %\end{small}
%        \vspace{-0.5cm} % Journal version
        \vspace{-0.30cm} % ArXiv version
\end{center}
\end{table}

%% END OF FILE

\acresetall
%\pagebreak
\section{Conclusion}\label{sec:conclusion}

%In this paper we first analyzed the overall volume of \ac{res} in Germany and second show the calculation of day-ahead prices under the assumption that power feed-in of wind and \ac{pv} units has to be settled with assumed positive marginal costs,~i.e., $0-25$\euro/MWh (Sec.~\ref{sec:marCostLiterature}), at the day-ahead market.
While \ac{fit}-subsidized \ac{res} generated electricity contributed less than one-fifth of Germany's total electric load demand per year, it represents almost half of the traded day-ahead spot volume. This is no surprise as the \ac{epex} day-ahead market only has a market share of about two-fifth of Germany's total load demand in recent years (2011--2013). The day-ahead market is therefore largely dominated by subsidized \ac{res} feed-in, which in turn drives the dynamics of the day-ahead price clearing mechanism. The well-known merit-order effect of \ac{res} power feed-in,~i.e.~falling day-ahead spot price levels as well as highly fluctuating day-ahead prices, is thus amplified to a level that is in clear contrast with the physical situation in the German power system, in which \ac{res} feed-in still plays a minor, albeit rapidly rising role.

An obvious approach for mitigating the merit-order effect within the existing energy-only power market,~i.e., without changing the market mechanism towards a capacity market or else, is to increase the day-ahead spot markets trading volume to actually capture the majority of the load demand. This can be achieved by incorporating more of the electricity production that is currently sold bilaterally,~i.e., \ac{otc}, outside of the day-ahead market. Comparably mature power markets like \emph{Nord Pool Spot} in the Nordic countries achieve a market share of more than four-fifth of the respective total load demand.
The authors believe that reducing the volume dominance of \ac{res} feed-in in the \ac{epex} spot by means of mandatory spot pools will significantly reduce the existing \ac{res}~merit-order effect. %The quantitative effect of the \ac{res} feed-in share at the day-ahead market on the merit-order effect is part of future investigations.

We also show that the settlement of \ac{res} at the day-ahead market with marginal cost is possible. Negative prices disappear and in most days, at least some of the \ac{res} production volume is settled. Base load power plants profit from the situation because of an elevated average day-ahead price level. Gas-fired power plants and hydro storage units covering peak hours are not effected, since peak prices are not effected by the above discussed changes. Because of the further decreasing peak-base spot price spread, the already non-profitable short term storage plants suffer even more. Given fully dispatchable power plants that have sufficient operational flexibility in technical terms, i.e.,~fast ramping, start/stop times and minimum operation point requirements, energy-only power markets seem to work even for high \ac{res} penetration scenarios~($\gg 25\%$).
% Journal version
%Future work will address the impact of such a regime change, which gives no \ac{res} feed-in guarantee, on the effective \ac{res} grid integration. 

%\begin{enumerate}
%    \item[?] Discussion Energy Security (fully dispatchable backup capacity is needed for power production in 	'emergency situations'. Questions are: [1] How much capacity [in \% of peak load, in GW] is necessary? 	[2] How can these backup capacities be made and kept economic [capacity market, payments for reserve 		provision, ...])?
%\end{enumerate}

%\appendices

%\section{}


%\section*{Acknowledgment}
%\input{Acknowledgment/Acknowledgment}

%\ifCLASSOPTIONcaptionsoff
%  \newpage
%\fi

\bibliographystyle{ieeetran}
%\bibliography{bib/endNoteLibrary_MH}
% \bibliography{bib/MH_Mend_Bib}
\bibliography{MeritOrderPaper,Biblio/PowerNodesBib,Biblio/GM2012,Biblio/IFAC_MPC,Biblio/RINKE_2011_MPC_Grid_Inertia,Biblio/diss,Biblio/endNoteLibrary_MH,Biblio/MeritOrderPaper,Biblio/PSCC,Biblio/Inertia_FullPaper,Biblio/EEM10_ETH_Zurich_ULBIG,Biblio/endNoteLibrary_EEM2011,Biblio/DLR_biblio}

\end{document}